# Zonally opposing shifts of the intertropical convergence zone in response to climate change

by

Antonios Mamalakis[1*], James T. Randerson[2], Jin-Yi Yu[2], Michael S. Pritchard[2], Gudrun Magnusdottir[2],

Padhraic Smyth[3,4], Paul A. Levine[2], Sungduk Yu[5], and Efi Foufoula-Georgiou[1,2*]

[1] Department of Civil and Environmental Engineering, University of California, Irvine
[2] Department of Earth System Science, University of California, Irvine
[3] Department of Computer Science, University of California, Irvine
[4] Department of Statistics, University of California, Irvine
[5] Department of Geology and Geophysics, Yale University, New Haven, Connecticut

*email: efi@uci.edu, amamalak@uci.edu

**Abstract**

Future changes in the location of the intertropical convergence zone (ITCZ) due to climate change are of high interest since they could substantially alter precipitation patterns in the tropics and subtropics. Although models predict a future narrowing of the ITCZ during the 21$^{st}$ century in response to climate warming, uncertainties remain large regarding its future position, with most past work focusing on the zonal-mean ITCZ shifts. Here we use projections from 27 state-of-the-art climate models (CMIP6) to investigate future changes in ITCZ location as a function of longitude and season, in response to climate warming. We document a robust zonally opposing response of the ITCZ, with a northward shift over eastern Africa and the Indian Ocean, and a southward shift in the eastern Pacific and Atlantic Ocean by 2100, for the SSP3-7.0 scenario. Using a two-dimensional energetics framework, we find that the revealed ITCZ response is consistent with future changes in the divergent atmospheric energy transport over the tropics, and sector-mean shifts of the energy flux equator (EFE). The changes in the EFE appear to be the result of zonally opposing imbalances in the hemispheric atmospheric heating over the two sectors, consisting of increases in atmospheric heating over Eurasia and cooling over the Southern Ocean, which contrast with atmospheric cooling over the North Atlantic Ocean due to a model-projected weakening of the Atlantic meridional overturning circulation.

1. **Introduction**

The intertropical convergence zone (ITCZ) and its dynamics[1] play a vital role in the tropical atmospheric circulation and hydroclimate, sustaining tropical forest and savanna ecosystems, and regulating the food security and property of billions of people. As such, intense research has been focused on identifying the physical mechanisms that determine the climatology and variability of the ITCZ position on intra-seasonal to interannual scales[1-10], and its long-term response to large scale natural climate variability and anthropogenic forcing[1,5,11-22]. Evidently, understanding the mechanisms that regulate the position of the ITCZ has been highlighted as an important knowledge gap in future global and regional climate change[23].

From a local perspective, the ITCZ position has been shown to be controlled by tropical mechanisms impacting near-equatorial sea surface temperature (SST) gradients[24]. From an energetics perspective, analysis of paleoclimate records, reanalysis data, and idealized climate simulations all indicate that the ITCZ variability can be influenced by differences in atmospheric heating between the northern and southern hemispheres (in the zonal mean, such hemispheric energy asymmetries determine the cross-equator atmospheric energy transport; $AET_0$, the zero in the subscript refers to the equator), with the ITCZ tending to shift toward the more heated hemisphere, mimicking its seasonal behavior[1]. Hemispheric energy asymmetries can be the result of natural climate variability (shifts in the Atlantic Multi-decadal Oscillation[25], or volcanic eruptions[26]) or anthropogenic forcing (e.g., changes in emissions of sulfate aerosols[16,19,27]). Many studies have also highlighted the importance of extratropical energy sources/disturbances in altering tropical dynamics[28-36]. In the case of a northern hemisphere atmospheric cooling, the ITCZ is displaced southward, increasing the northward $AET_0$ to maintain the atmospheric energy balance[31] (a recent example of a southward ITCZ shift occurred during the late 20$^{th}$ century[37], likely because of increasing emissions of sulfate aerosols in the northern hemisphere, which decreased the temperature difference between the northern and southern hemispheres[16,19,21,38]). Apart from $AET_0$, ITCZ variations have been also linked to the equatorial net energy input into the atmosphere ($NEI_0$), with the ITCZ shifting equatorward when $NEI_0$ increases (e.g., during an El Niño event)[6,9]. More generally, the ITCZ has been shown to covary with the energy flux equator (EFE; a zone where the meridional AET vanishes[31]), the latitude of which can be, to a first order, approximated by the ratio of $AET_0$ and $NEI_0$[1,6,9]; a proxy that combines in a sense both local (mostly reflected in $NEI_0$) and non-local (reflected in $AET_0$) energy sources/disturbances. The close link between the location of the ITCZ and the EFE holds not only in the zonal mean[1,9], but also over large longitudinal sectors such as a continent or ocean basin[10], which has motivated recent studies to try to explain sector-mean ITCZ variability using a "two-dimensional (2D) energetics framework"[10,39-42]. Under a 2D energetics framework, both zonal and meridional fluxes are taken into account, allowing exploration of how regional variation in ITCZ location may be linked with longitudinal differences in extratropical processes.

Regarding the response of the ITCZ to future climate change in particular, past research has mostly focused on zonal-mean changes, and it shows that many different factors (greenhouse gases, aerosols, albedo, clouds, ocean heat transport or storage, and regional ocean circulation) can affect the geographic pattern of tropical SSTs and/or the energy balance, and consequently the ITCZ location[18,22,32]. For example, future reduction in aerosol emissions[27,38,43], as well as Arctic sea-ice loss (related to Arctic amplification[44,45]) and glacier melting in the Himalayas[46,47] are expected to reduce albedo significantly more in the northern hemisphere than in the southern hemisphere, resulting in a northern warming and an ITCZ shift to the north[18,22,34]. In contrast, the Atlantic Meridional Overturning Circulation (AMOC) is expected to weaken in the future[48-51] (new results indicate that it has already been weakening[52]), which will result in a reduction of the northward oceanic heat transport from the tropics to the northern Atlantic and a northern cooling, leading to a southward shift of the ITCZ[22,29,34,53].

Despite the relative consensus in the literature with regard to the zonal-mean response of the ITCZ location to individual forcing agents and processes as discussed above, there is still uncertainty (i.e. large inter-model spread) about the response of the ITCZ location to the integrated effect of all these processes under climate change. This uncertainty mainly stems from different model physics that yield different responses even to identical forcing from a single representative concentration pathway. Particularly, although a future narrowing of the ITCZ is a robust projection expected with climate change[20], models differ considerably regarding changes in the position of the ITCZ, yielding to an almost zero zonal-mean ITCZ shift when considering the multi-model mean[22]. Another reason for this uncertainty is that, as mentioned earlier, most studies have focused on zonal-mean changes of the ITCZ, possibly masking model agreements over particular areas. Indeed, because of the compensating effects of the relevant radiative and dynamical processes influencing the ITCZ position (e.g. the northward ITCZ shift caused by snow and ice albedo feedbacks and reduction of aerosols in the northern hemisphere will be compensated by a weaker AMOC[22]), and since most of these processes are not expected to be equally influential in different longitudinal sectors of the globe, the integrated ITCZ response to climate change should not be expected to be homogeneous in longitude[18]. Thus, more explicit analysis, focusing on the regional ITCZ and EFE changes (rather than zonal-mean changes) is necessary to gain insight into the future response of the ITCZ location to climate change, and to identify robust model projections across different longitudinal sectors.

In the light of the above, here we explore the ITCZ responses to climate change from the present through 2100 using Earth system model simulations from the sixth phase of the Coupled Model Intercomparison Project[54] (CMIP6; see Supplementary Table 1) forced with the SSP3/RCP7.0 scenario[55,56] (that is, the combination of the shared socioeconomic pathway 3 and the representative concentration pathway 7.0). In our analysis, we explicitly examine seasonal and annual-mean ITCZ position changes as a function of longitude, while also taking into account the present-day ITCZ biases of each model. In this way, we aim to identify model consensus over particular areas, elucidate the regional responses of the ITCZ to climate change, and gain insight into the processes that influence the ITCZ location in different longitudinal sectors.

With regard to regionally tracking the ITCZ, ambiguity exists in the literature as to a precise regional definition of the ITCZ and/or which is the optimal variable/method to use for tracking its position[57]. For example, past studies have variously used surface pressure minimum, surface wind convergence, precipitation maximum, minimum outgoing longwave radiation (OLR) or cloudiness maximum to track the ITCZ[57]. The justification for using so many different variables to track the ITCZ is the assumption that the minima or maxima of these different variables collocate with each other (i.e. pressure minima roughly collocate with convergence maxima, etc.); yet, this assumption may not be true over specific regions or in specific seasons[57], and so, this ambiguity in the regional ITCZ definition is problematic. For the purpose of this study, to address the latter ambiguity, we have used a multivariate probabilistic framework[37], which tracks the ITCZ over different longitudes and seasons by simultaneously assessing the statistics of multiple variables, and thus increasing the robustness of the tracking approach (see reference [58] for other probabilistic methods which use multiple variables to track the ITCZ). In particular, we consider overlapping longitudinal windows, and use the window-mean precipitation and OLR (the two most common variables in the ITCZ literature) to track the ITCZ. For each window and season, ITCZ points are defined as those which correspond to the maximum (above a certain threshold) joint probability of non-exceedance of the two window-mean variables (note that in cases where precipitation and OLR extrema collocate, the latter definition falls back to simply tracking the points of the extrema, and results would be identical if we were to use either variable on its own; see section *Methods*, and [37] for more information). The end product of the method is to provide the probability of every grid point in the tropics to be part of the ITCZ in a longitudinally-explicit manner (see Supplementary Figure 1). The resulting probability distribution of ITCZ position is used to compare the climatology and interannual variability of the ITCZ between observations and CMIP6 models during a contemporary base period (1983-2005), as well as to assess future ITCZ changes (defined as the difference between 2075-2100 and the base period). To complement this view, in specific cases, we also present results based on simple univariate precipitation or OLR maps/indices to assess ITCZ changes. This is for the sake of completeness and to demonstrate that our inferred trends in ITCZ position are robust with respect to the tracking methodology.

In our analysis, we use satellite data of precipitation[59] and OLR[60] as our reference datasets for building an ITCZ position climatology during the base period, and simulations from 27 different CMIP6 models (a total of 105 individual runs when including initial condition ensembles) to explore the effect of climate change on ITCZ location by 2100; see Supplementary Table 1 and section *Methods* for more information. The ITCZ shifts we identify are then evaluated for physical consistency with future changes in equatorial SSTs, $AET_0$, and EFE shifts.

2. Results

**Model simulation of contemporary ITCZ position.** Model simulations of the historical ITCZ climatology are known to exhibit important biases (e.g. the so-called "double-ITCZ biases" [61,62]). Thus, estimates of future ITCZ shifts, which are obtained as the difference between the simulated future and baseline averages, need to be cautiously interpreted and analyzed. Particularly, including information about the present-day ITCZ model biases in the analysis may lead to a better understanding of future ITCZ shifts, as recent literature suggests [63,64]. In terms of the climatological mean location of the ITCZ, we find that although the models are mostly consistent in simulating the location of the ITCZ during May-Oct, they exhibit important double-ITCZ biases in the Eastern Pacific and Atlantic Oceans in the season Nov-Apr (see Figure 1), which have been well documented and explored as to their linkage with other systematic biases in simulated equatorial SSTs and the atmospheric energy input/transport [33,40,62,65-67]. In order to assess the impact of these present-day model biases on our interpretation of the future ITCZ trends more quantitatively, we calculated the ITCZ biases for each model and over each basin, by obtaining the spatial average of the difference in the (Nov-Apr) probability distribution of the ITCZ location between models and observations over specific boxes (see Figure 1d, Supplementary Figure 2a and section *Methods* for more information). Our results indicate that CMIP6 models generally simulate a more frequent southward migration of the Atlantic ITCZ than what is observed, by $\Delta P = 57 \pm 17.8\%$ (that is the difference in probability between models and observations), and likewise in the Pacific toward the southeastern sector of the basin, by $\Delta P = 34 \pm 11.3\%$ (see Supplementary Figure 2b and Supplementary Table 1). These numbers show that the Atlantic bias has a larger magnitude, and as such, the signature of the seasonal double-ITCZ biases on the annual scale is apparent mainly over the Atlantic and to a lesser degree over the eastern Pacific basin (see Figure 1f). Note that when we use the average tropical precipitation or OLR difference between models and observations to assess the systematic double-ITCZ biases (i.e. not using the probabilistic method but using a traditional approach), we obtain similar results (see Supplementary Figure 3). In terms of the interannual variability of the ITCZ, CMIP6 models are found to correctly produce the expected equatorward shift of the ITCZ during El Niño events (Supplementary Figure 4). For more information on the observed ITCZ climatology and model biases, see Supplementary Discussion.

# Distribution of the location of the ITCZ during 1983-2005

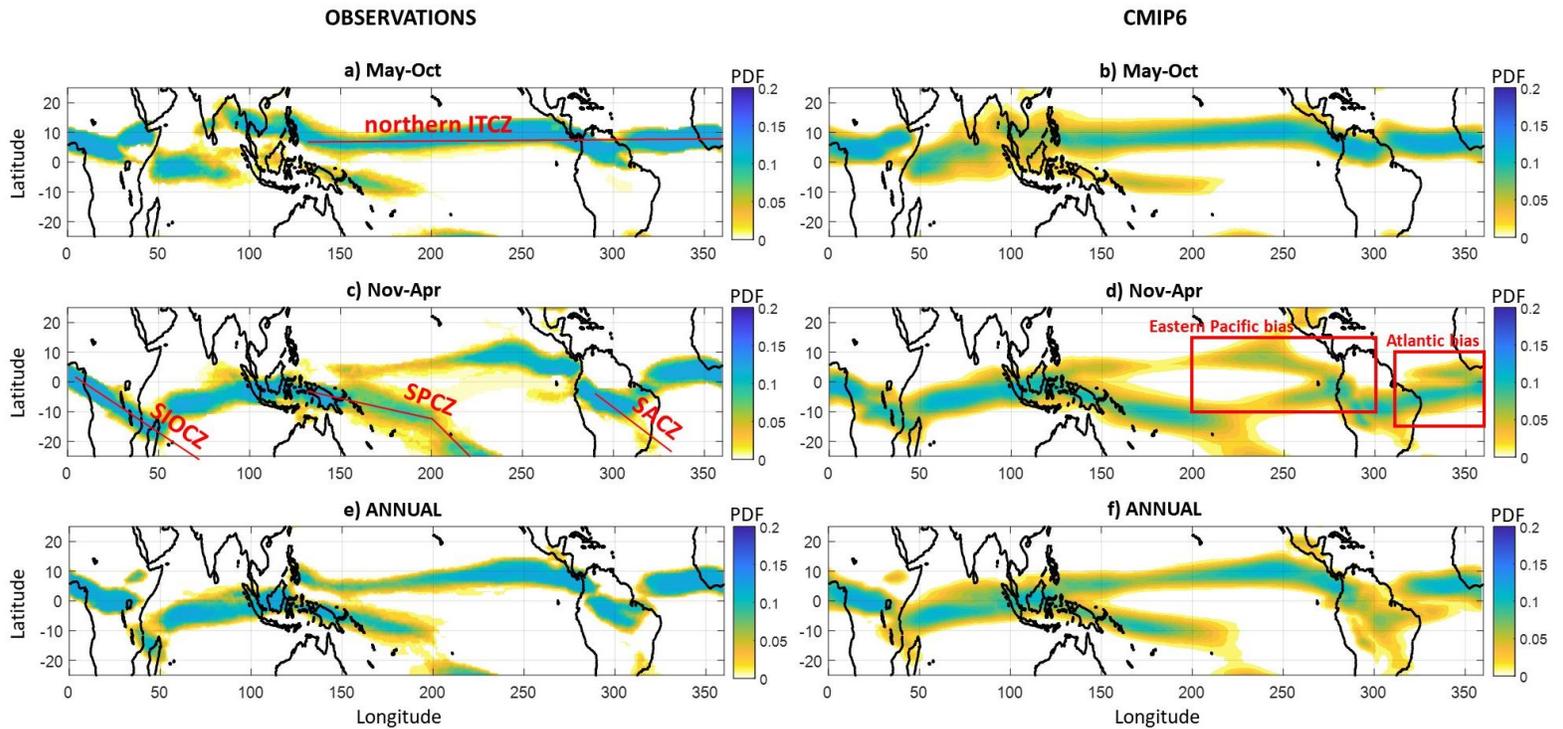

**Figure 1: Climatological location of the ITCZ during 1983-2005 based on observations and CMIP6 models.** a-b) Probability density function (PDF) of the location of the ITCZ in all longitudes during season May-Oct. The ITCZ tracking is performed based on the joint statistics of the observed (panel (a)) or the simulated (panel (b)) window-mean precipitation and outgoing longwave radiation (OLR) in overlapping longitudinal windows (see Supplementary Figure 1 and section *Methods*). c-d) Same as in (a)-(b), but for season Nov-Apr. e-f) Same as in (a)-(b), but tracking is performed on an annual scale. In (b), (d), and (f), the multi-model mean across all 27 CMIP6 models is presented. Some well-known distinct ITCZ features are highlighted in the results from the observations (see Supplementary Discussion), while the double-ITCZ biases in the eastern Pacific and Atlantic basins are apparent in the CMIP6 results (season Nov-Apr). The areas over which the double-ITCZ biases are quantified are shown as red boxes in panel (d); see section *Methods* for more information.

**Future ITCZ change.** Our analysis of annual and zonal-mean changes indicates that the ITCZ shift under future climate change for the CMIP6 models is -0.5 ±1.2° N (slightly southward; see Table 1). The inter-model uncertainty within the CMIP6 models is very large (the standard deviation is more than twice the mean shift), which leads to the multi-model mean shift not being statistically distinguishable from zero (Table 1), and confirming previous reports[22].

One of the important findings of this study is that despite the high inter-model uncertainty regarding the zonal-mean ITCZ shift, models exhibit greater agreement in ITCZ changes as a function of longitude (see Figure 2). Particularly, in the May-Oct season, CMIP6 models indicate a robust northward shift of the ITCZ over Africa and the Indian Ocean[68], and a southward shift over most of the Pacific and Atlantic Oceans (in the Atlantic basin the shift is less statistically robust relative to the one in the Pacific; see Figure 2a). In the Nov-Apr season, the south Indian Ocean convergence zone and the south Pacific convergence zone both shift northward[63], while the eastern Pacific ITCZ is shown to shift southward. In the Atlantic basin, most models predict a higher probability of the ITCZ to prevail over the equator in the future relative to the base period, revealing a pattern of an equatorward ITCZ shift. In general, an interesting zonally opposing ITCZ response to climate change is revealed in both seasons and even more clearly on the annual scale (see Figure 2c and Table 1), which consists of a robust northward ITCZ shift over the eastern Africa and Indian Ocean region, and a robust southward ITCZ shift over the eastern Pacific Ocean, South America, and the Atlantic Ocean. This zonally opposing response is also apparent when calculating the future change in annual-mean precipitation or OLR (see Supplementary Figure 5).

To more precisely quantify this zonally opposing ITCZ response to climate change, we tracked the temporal evolution of the ITCZ location as a function of longitude and over two different longitudinal sectors, i.e. the Eurasian sector (20°E-130°E) and the eastern Pacific and Atlantic sector (250°E-360°E); the boundaries of the two sectors were chosen based on Figure 3a, but our results are quite robust if the boundaries are moderately changed (i.e., by ±10°). A clear northward ITCZ shift is observed over the Eurasian sector, while a southward shift is apparent in the eastern Pacific and Atlantic Oceans (Figure 3a). Over the western Pacific, the ITCZ shifts southward during May-Oct and northward during Nov-Apr (as was shown in Figure 2), which translates into a decreased seasonal ITCZ migration in the future, and an annual-mean shift that is close to zero. When comparing the 2075-2100 and 1983-2005 periods, a statistically significant (based on the *t*-test; $p < 0.01$) northward shift on the order of $0.8 \pm 0.6°$ N is obtained over the Eurasian sector (see Table 1). In contrast, over the eastern Pacific and Atlantic sector, CMIP6 models indicate a statistically significant southward shift on the order of $-0.7 \pm 0.9°$ N. The future ITCZ shift and the corresponding change in annual-mean precipitation asymmetry (i.e. the change in the quantity: Precip $_{0°-20°N}$ − Precip $_{0°-20°S}$) between 2075-2100 and 1983-2005 are shown for every CMIP6 model in Figure 3b, indicating that the majority of models predict a future increase in precipitation in the northern subtropics relative to the south over the Eurasian sector (red color). The opposite is true for most CMIP6 models over the east Pacific and Atlantic sector (blue color). The latter results also illustrate the robustness of the revealed zonally opposing ITCZ response with respect to using different indicators to assess ITCZ changes (i.e. precipitation asymmetry vs the probabilistic tracking).

With regard to the effect of the double-ITCZ biases on the revealed ITCZ response, we find that the results over the Eurasian sector are not sensitive to the performance of the models in the base period. That is, there is no statistically significant relationship between the double-ITCZ biases and the projected shift over the Eurasian sector across CMIP6 models (not shown). However, over the eastern Pacific and Atlantic sector (i.e. where the double-ITCZ biases occur), the double-ITCZ biases seem to influence the sign of the predicted ITCZ shift to some extent. In particular, our analysis shows that the smaller the bias of a model over the southern Atlantic, the more likely it is to predict a southward shift of the Atlantic ITCZ in the future (see Supplementary Figure 6). This implies that the pattern of the ITCZ contraction over the Atlantic Ocean that is depicted in e.g. Figure 2b is likely a spurious result, originating from some of the models being highly biased during the base period. Since in reality the Atlantic ITCZ remains in the northern hemisphere for most of the year and there is very little to zero precipitation over the southern Atlantic Ocean (see Figure 1 and Supplementary Figure 3), a future southward Atlantic ITCZ shift as indicated by the models with lower bias is more likely (a future negative ITCZ-pattern over the southern Atlantic Ocean as shown in Figure 2b is an artifact from the high bias in some models and will be an algebraic impossibility in reality). In support of this interpretation, we find that the small number of CMIP6 models that predict a northward ITCZ shift over the east Pacific and Atlantic sector (i.e. in contrast to the majority of the models that predict a southward ITCZ shift; see Figure 3b) exhibit relatively high double-ITCZ biases in the base period. Thus, we argue that the double-ITCZ biases, if anything, are obscuring the full extent of the southward ITCZ shift over the eastern Pacific and Atlantic sector, and thus, our result of the zonally opposing response of the ITCZ to climate change is on the conservative side.

Overall, the robust agreement between CMIP6 models over these two large sectors (Table 1 and Figures 2-3) provides confidence that climate change will lead to contrasting meridional shifts of the ITCZ in the Eurasian vs. E. Pacific/Atlantic sectors. As already mentioned, these contrasting responses in the different longitudinal sectors nearly cancel one another, leading to almost zero ITCZ shift from a zonal-mean perspective (Table 1), confirming the recent literature[22,69].

**Table 1:** Mean and standard deviation of the future ITCZ and EFE shifts (2075-2100 minus 1983-2005, positive values indicate northward movement) and changes of the hemispheric energetic asymmetry over different longitudinal sectors, as obtained from 27 CMIP6 model outputs. The baseline values (i.e. referring to 1983-2005) are also provided. Values with **bold** font correspond to a multi-model mean which is statistically distinguishable from zero, based on the *t*-test ($p < 0.01$). It is shown that there is a robust consensus across models regarding future changes in the Eurasian and E. Pacific – Atlantic sectors, but such a consensus is not apparent in the zonal mean. Note for example that in the sector-mean analysis, the inter-model variability (i.e. st. deviation) in future changes is either smaller or of the same magnitude with the multi-model mean, while in the global zonal-mean analysis the inter-model variability is in all cases 2 to 4 times larger than the multi-model mean.

| 27 CMIP6 Models | | Global zonal mean | Eurasian Sector [20ºE-130ºE] | E Pacific & Atlantic Sector [250ºE-360ºE] |
|---|---|---|---|---|
| ITCZ latitude (degrees North) | Base Period | **3.6 ±2.0** | **-1.0 ±1.1** | **4.1 ±2.3** |
| | Future Shift | -0.5 ±1.2 | **0.8 ±0.6** | **-0.7 ±0.9** |
| $Q_S - Q_N$ (PW) | Base Period | -0.03 ±0.37 | **0.93 ±0.21** | **-0.96 ±0.23** |
| | Future Change | -0.05 ±0.21 | **-0.24 ±0.10** | **0.31 ±0.16** |
| EFE latitude (degrees North) | Base Period | -0.4 ±0.8 | **-3.2 ±0.9** | **4.9 ±1.9** |
| | Future Shift | 0.2 ±0.5 | **0.7 ±0.4** | **-1.4 ±1.1** |

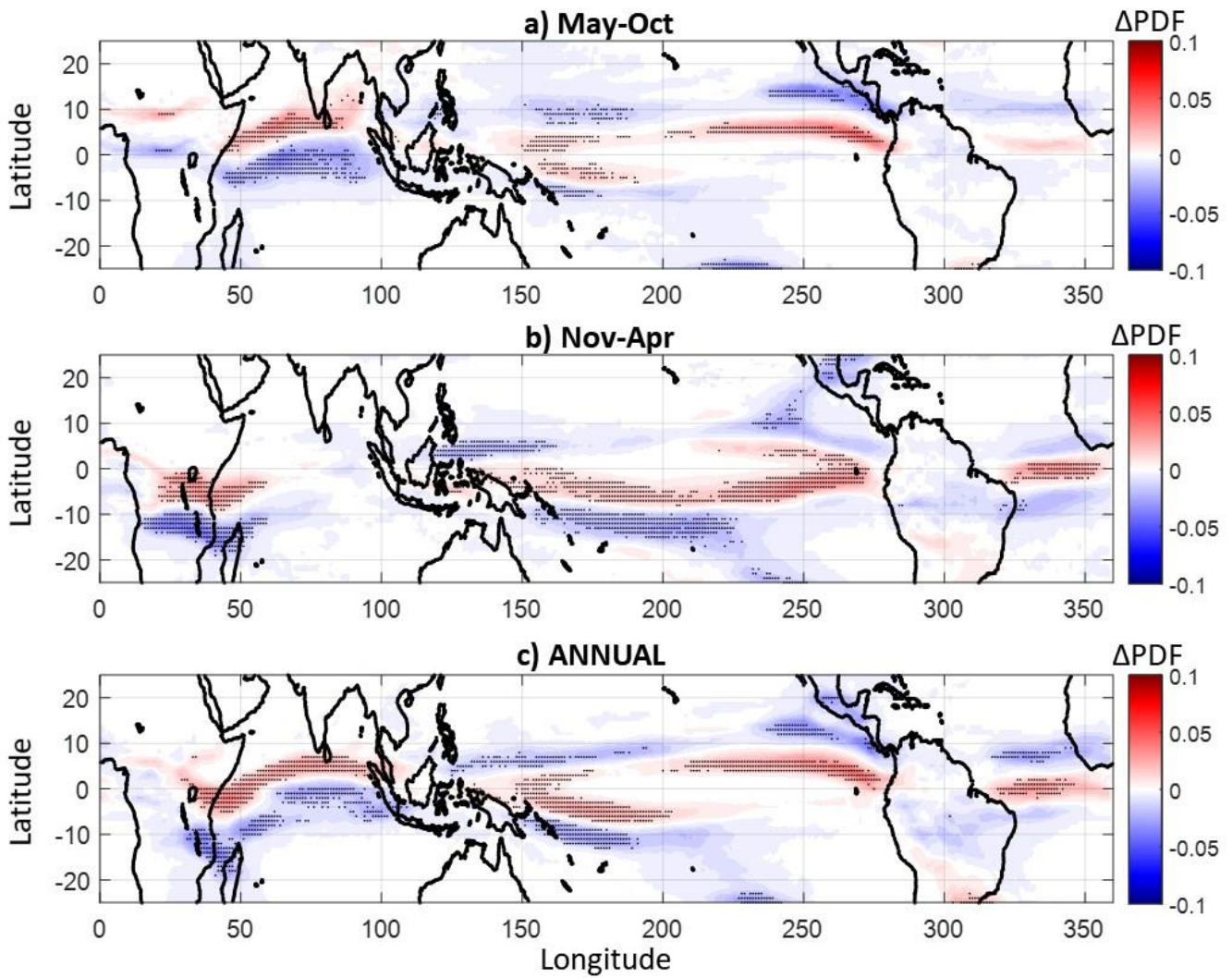

**Figure 2: Future changes in the ITCZ location in response to climate change, as predicted by CMIP6 models.** a) Difference in the probability density function (ΔPDF) of the location of the ITCZ in May-Oct between 2075-2100 and 1983-2005. b) Same as in (a), but for Nov-Apr. c) Same as in (a), but the changes in the annual distribution are shown. In all plots, the multi-model mean across 27 CMIP6 models is presented, while stippling indicates agreement (in the sign of the change) in more than ¾ of the models considered. Results indicate a robust northward ITCZ shift over eastern Africa and Indian Ocean and a southward ITCZ shift over eastern Pacific and Atlantic Oceans.

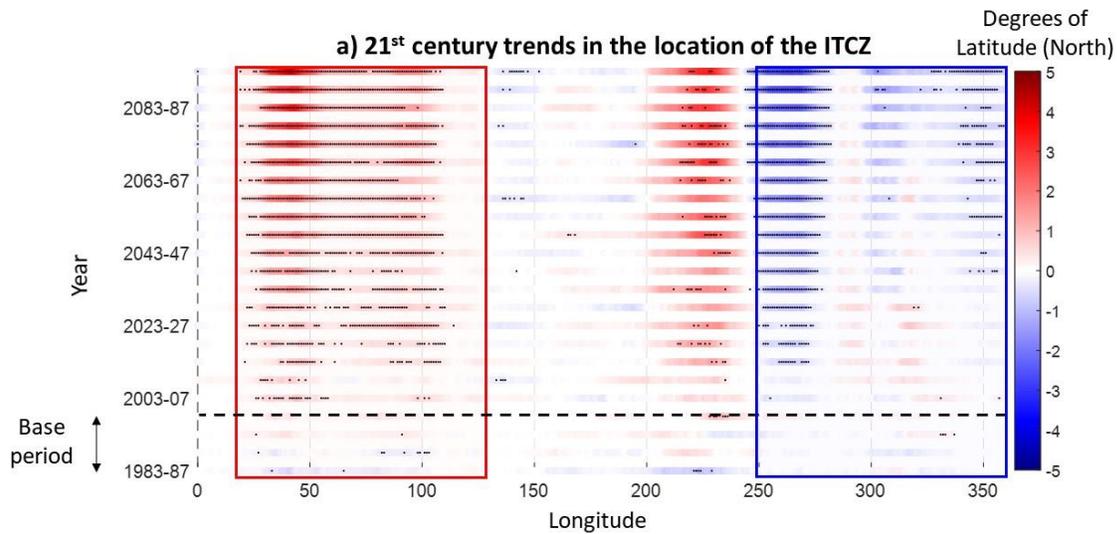

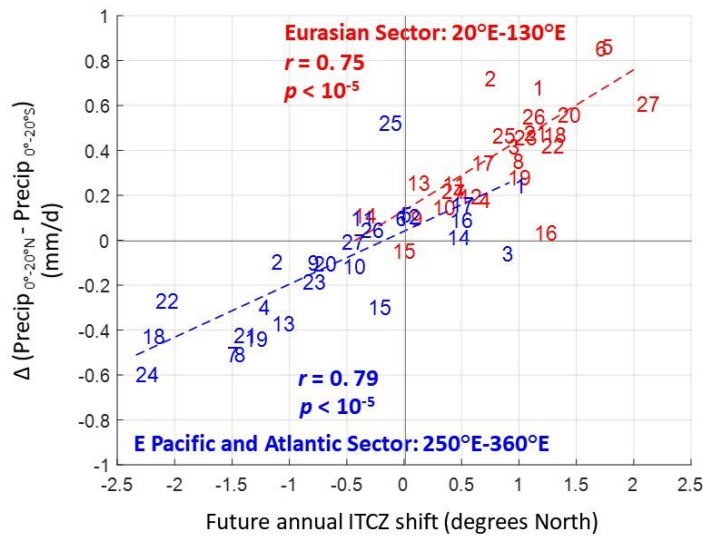

**Figure 3: 21st century series of ITCZ location as predicted by CMIP6 models.** a) Series of the 5yr-mean ITCZ location relative to the base period as a function of longitude. The multi-model mean across 27 CMIP6 models is presented, while stippling indicates agreement (in the sign of the change) in more than ¾ of the models considered. b) Scatter plot of the projected ITCZ shift (horizontal axis) and change of tropical precipitation asymmetry (vertical axis) between the periods 2075-2100 and 1983-2005, using all 27 CMIP6 models zonally averaged over the Eurasian sector (20ºE-130ºE; red color) and the eastern Pacific and Atlantic sector (250ºE-360ºE; blue color). Each model is labeled according to Supplementary Table 1. Based on either index (ITCZ shift or precipitation asymmetry), a robust opposing ITCZ response between the two sectors is revealed, whereby the ITCZ is projected to shift northward over the Eurasian sector and southward over eastern Pacific and Atlantic Oceans.

**Link to sea surface temperature changes.** Many different tropical explanations/mechanisms contributing to future and past regional ITCZ and precipitation shifts have been proposed in the literature (e.g. the wet-get-wetter mechanism[70], feedbacks affecting near-equatorial sea surface temperatures[68], plant physiological responses[71], changes in monsoonal dynamics[72], etc.). Motivated by the known close coupling between sea surface temperature and precipitation in the tropics[68,73,74], we explored the consistency of the revealed zonally opposing shifts of the ITCZ with changes in the SST. We find that, globally, SST warming is more pronounced in the northern hemisphere than the southern hemisphere (Figure 4). This is a known and robust result under climate change and is partially due to the strengthening of the southeast trade winds which favor sea surface evaporation[68]. Another important contributor to this hemispheric asymmetry is the muted warming in the Southern Ocean (Figure 4), which has been attributed to the intense vertical mixing that occurs in this area, resulting in considerable ocean heat uptake[75]. In fact, models and observations suggest that more than half of the historical excess heat due to the increased radiative forcing has been stored in the Southern Ocean over the last decades[75-77]. Muted warming is also observed over the north Atlantic Ocean, which is likely to be a result of the weakening of the AMOC, another robust feature under climate warming[50,51].

Regarding the tropics, we find that over the Pacific Ocean, SST warming is more pronounced in the east than the west, which is a consistent result with the anticipated weakening of the Walker circulation under climate change[68,78]. In both the eastern Pacific and Atlantic Oceans, higher SST warming occurs at low latitudes between 10°S and 5°N, which is consistent with this region serving as an attractor for a southward shift of the ITCZ from its current baseline position at 4.1 ± 2.3°N for this sector (see Figure 4c). In contrast, over the Indian Ocean in the Eurasian sector, higher SST warming in the northern subtropics is consistent with the predicted shift of the ITCZ to the north from its current baseline position (Figure 4b). The pattern of SST change in the Indian Ocean resembles a positive Indian Ocean Dipole (IOD) pattern (with more pronounced warming over the northwestern Indian Ocean and less pronounced warming over southeastern Indian Ocean), traditionally linked to locally developed Bjerknes feedbacks between SST gradients, and wind and thermocline changes in the basin[68,78,79].

Despite the fact that the predicted changes in tropical north-south SST gradients are consistent with the zonally opposing ITCZ response, more insight is needed as to why these SST changes occur. Both local and non-local process chains are relevant. For example, the positive IOD pattern in the Indian Ocean has been argued to be a result of the weakening of the Walker circulation locally, but also influenced at its southern margin by the oceanic lateral advection of relatively weak warming signatures from the remote Southern Ocean[68]. Other causatively relevant non-local possibilities include extratropics-to-tropics teleconnections, which are usually based on energetic arguments[31,68]. Indeed, as noted in the introduction, to get more insight into past or future ITCZ shifts, recent studies have utilized atmospheric energetic constraints to explain tropical climatic changes, and in some cases have attributed them to extratropical factors, even if these changes were longitudinally confined, i.e. not referring to the zonal mean[10,39-42,53]. Motivated by this, we looked into the future changes in the atmospheric heat budget and further investigated whether the zonally opposing ITCZ response could be related to similar zonally opposing changes in the hemispheric heating and EFE shifts.

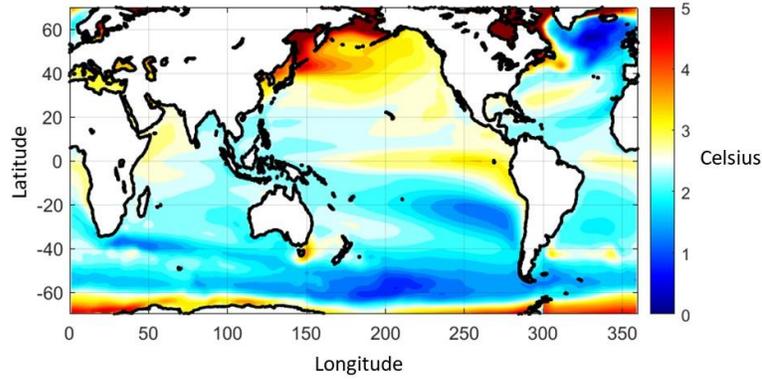

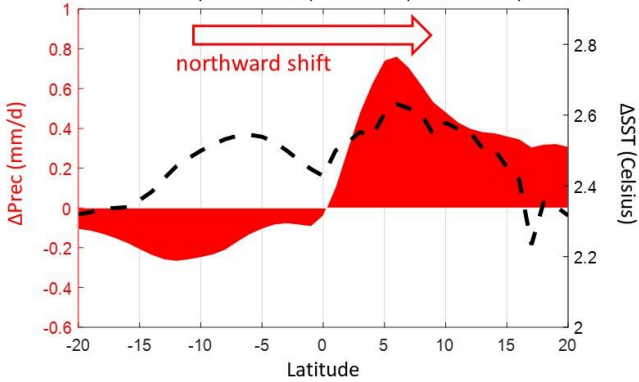 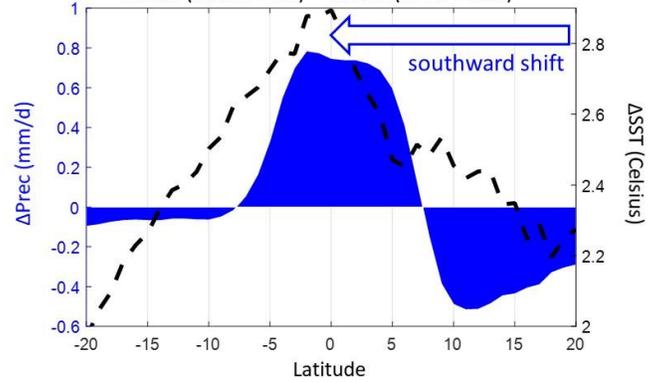

**Figure 4: Future changes in sea surface temperature and precipitation in response to climate change, as predicted by CMIP6 models.** a) Global changes in sea surface temperature (SST) between future 2075-2100 and base period 1983-2005. b) Zonal mean over the Indian Ocean (50ºE-100ºE) of the changes of precipitation (in mm/d) and SST (in Celsius). c) Same as in (b), but for the eastern Pacific and Atlantic Oceans (250ºE-360ºE); land changes are not considered in the zonal mean. All results refer to the multi-model mean across 27 CMIP6 models.

**Atmospheric Energetic constraints.** Considering a long enough period (e.g. 1983-2005) so that the energy storage in the atmosphere is negligible[1,80], and assuming that the system is in equilibrium, the atmospheric energy budget is [6,80]:

$$\nabla \cdot \mathbf{F} = R^{\text{TOA}} - O = Q \qquad (1)$$

where **F** is the vector of vertically-integrated atmospheric moist static energy flux, $R^{\text{TOA}}$ is the net energy input at the top of the atmosphere (TOA; i.e. net downward shortwave minus the outgoing longwave radiation) and $O$ is the ocean energy uptake (can be further partitioned to latent/sensible heat and radiative surface components) and represents the heating from the surface (note that the energy storage in the land is negligible on timescales greater than a season[1]). $Q$ is the net energy input into the atmospheric column of unit horizontal area (see schematic in Figure 5a, and Supplementary 7a-b for the distribution of $Q$ in the base period), and Equation (1) states that it is equal to the horizontal divergence of the AET. Thus, future changes in $Q$ as a response to climate change are related to changes in the horizontal AET, which can in principle be related to the future shifts in the ITCZ.

In response to climate change, models indicate that the energy input into the atmosphere will increase in the tropics and decrease at high latitudes 50º-70º, especially over the ocean (see Figure 5b for the change in the total energy input, and its partitioning into TOA and surface components[81] in Figure 5c-d, according to Equation (1)). Particularly, over the Atlantic Ocean, a pattern of northern atmospheric cooling and southern heating is revealed, which is consistent with a weakening in the AMOC (i.e. the see-saw response[22,29,50,51,82-84]), while over the Southern Ocean, atmospheric cooling is consistent with increased heat flux from the atmosphere to the ocean in response to increasing emissions of greenhouse gases[85]. Moreover, we find an increase in atmospheric heating over the tropics, which is mostly a result of the TOA component of the budget (Figure 5c), and is likely associated with water vapor and cloud radiative effect; i.e., the OLR escaping to space is reduced in the future (see partitioning of TOA energy change in Supplementary Figure 8c, and [22]). Over land, the effect of processes like snow and ice albedo feedbacks (see Supplementary Figure 8d and studies regarding climate change-induced glacier melting over the Himalayas[46,47], climate change-induced sea ice loss in the Arctic[34,44,86]) and reduction of anthropogenic aerosols, which are more pronounced over the northern hemisphere[22,43], are partially compensated by increases in OLR cooling (see Supplementary Figure 8b). As a result, we find that the net effect of all these processes is that more energy is being added into the atmosphere over land in the northern hemisphere and specifically over Europe, Southeast Asia, North America, and the Arctic (see Figure 5b), which contrasts the important heat loss in the northern Atlantic region due to the weakening of the AMOC.

In terms of the zonal mean, the compensating effects of all these processes lead to an almost zero net change in the hemispheric energy asymmetry. Particularly, CMIP6 models predict a change on the order of $\Delta(Q_S - Q_N)$ = -0.05 ± 0.21 PW ($Q_S$ and $Q_N$ refer to the hemisperically integrated atmospheric energy input over the southern and northern hemisphere, respectively) consistent with the negligible zonal-mean ITCZ shift (see Table 1). However, when considering the Eurasian sector and the eastern Pacific and Atlantic sector separately, significant differences emerge (Table 1), with models showing a higher level of consensus in terms of the sign of the change in the energy asymmetry (changes are assessed to be statistically significant; $p < 0.01$). Over the Eurasian sector, most models predict that more energy is added into the northern hemisphere than the southern hemisphere under climate change (Figure 5e), which reduces the baseline hemispheric energy asymmetry; i.e. $\Delta(Q_S - Q_N)$ = -0.24 ± 0.10 PW (see Table 1). In contrast, over the eastern Pacific and Atlantic Oceans, the northern hemisphere atmosphere receives less energy in the future (Figure 5e) probably due to the weakening of the AMOC, which contributes to a northern hemisphere atmospheric cooling; i.e. $\Delta(Q_S - Q_N)$ = 0.31 ± 0.16 PW. These results highlight opposing changes of the hemispheric energy asymmetry to global climate change between the two considered sectors, which is statistically consistent with the revealed zonally opposing response of the ITCZ (there is statistically significant dependence between changes in hemispheric heating and precipitation asymmetries; not shown), i.e. our results suggest that the ITCZ shifts towards the more heated hemisphere in each of the two sectors. However, such a suggestion is generally physically grounded only in the zonal mean perspective. Since these results do not refer to the zonal mean, more extensive analysis (e.g. considering the zonal energy fluxes at the sectors' boundaries as well as the $NEI_0$[1,6,9]) is needed in order to gain more insight into the quantitative link between future sector-mean ITCZ shifts and their regional energetic constraints. In doing so, we use a 2D theoretical energetics framework (where both zonal and meridional fluxes are taken into account)[10,39], which has only recently been used to explain sector-mean ITCZ shifts[10,39-42] and to the best of our knowledge, it has not yet been applied in any climate change scenario.

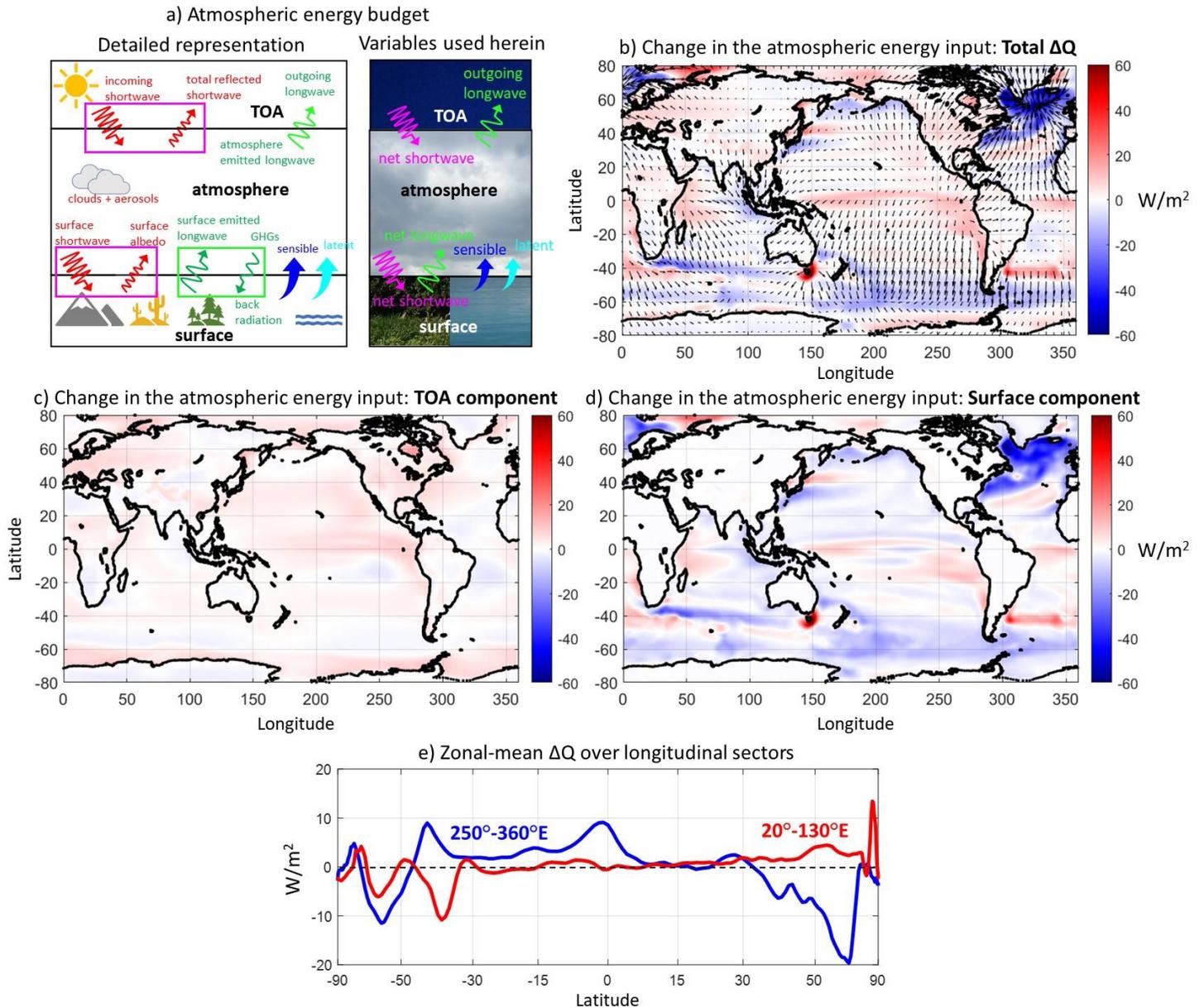

**Figure 5: Future changes in the atmospheric energy input in response to climate change, as predicted by CMIP6 models.** a) Graphic representation of the atmospheric energy budget. b) Difference of the average energy atmospheric input between 2075-2100 and 1983-2005 periods (shading), while vectors show the change in the divergent component of the atmospheric energy transport; vectors are on the order of $10^7$ W/m (see Figure 6 for specific values). c) Same as in (b), but only the top of the atmosphere (TOA) component is shown. d) Same as in (b), but only the surface component is shown. This panel highlights the contribution of the ocean to the future atmospheric heating/cooling. e) Zonal mean of (b) over the Eurasian sector (20ºE-130ºE; red curve) and the eastern Pacific and Atlantic sector (250ºE-360ºE; blue curve). The horizontal axis is scaled as $\sin(\varphi)$. In all plots, the multi-model mean across 27 CMIP6 models is presented. Results show that under global climate change, more energy is added in the atmosphere over the northern hemisphere than the southern hemisphere in the Eurasian sector, while the opposite is true in the eastern Pacific and Atlantic Oceans.

The energy flux **F** in Equation (1) can be decomposed into the divergent and rotational components (**F**$_\chi$ and **F**$_\psi$, respectively), and since the divergence of the rotational component is identically zero (i.e. $\nabla \cdot \mathbf{F}_\psi = 0$), Equation (1) takes the form of Poisson's equation:

$$\nabla \cdot \mathbf{F}_\chi = \nabla^2 \chi = Q \qquad (2)$$

where $\chi$ is the energy flux potential (an arbitrary scalar function)[10,39], such that its gradient is equal to the divergent component of AET, i.e. $(\partial_x \chi, \partial_y \chi) = \nabla \chi = \mathbf{F}_\chi = (u_\chi, v_\chi)$. By solving Equation (2), the potential $\chi$ (also $\mathbf{F}_\chi$) can be obtained; all derivatives are evaluated in spherical coordinates but written here in Cartesian coordinates for simplicity. In accordance to previous studies[10,39,40], outside from the tropics, the zonal component of the divergent AET is negligible compared to the meridional component in the base period (i.e. $v_\chi \gg u_\chi$; see Supplementary Figure 7c), while in the tropics, they are of the same magnitude (i.e. both the Walker and Hadley circulations contribute to the divergence of heat; Supplementary Figure 7d-e). The changes of $\mathbf{F}_\chi$ in response to climate change are presented in Figure 5b, and they are shown to be consistent with changes in $Q$. Noticeable features in these changes are the patterns of divergence over increased heating in the tropics and Europe, and the patterns of convergence over increased cooling of the atmosphere in the Southern and north Atlantic Oceans.

With regard to the changes of $\mathbf{F}_\chi$ specifically in the tropics (where the mean circulation, and thus, the ITCZ controls the AET), the results are insightful. Over the tropics of the Eurasian sector, a robust increase of southward energy transport is apparent in the future (see Figure 6a), which is consistent with the revealed northward shift of the ITCZ. In contrast, the future cooling over the Atlantic Ocean (Figure 5b) is compensated by changes in the extratropical $\mathbf{F}_\chi$ (likely controlled by extratropical eddies; see Figure 5b), but also by a robust increase in the northward energy transport over the tropics of the eastern Pacific and Atlantic (See Figure 6a), which is consistent with the revealed southward shift of the ITCZ in this sector. Similarly to the changes in the ITCZ location and in $Q$, these results highlight zonally opposing changes in the meridional component of $\mathbf{F}_\chi$, providing more confidence regarding the opposing ITCZ shifts over the two considered sectors. Note also that future changes in the zonal energy fluxes roughly resemble the opposite of the baseline pattern (i.e. opposite in sign and about 10% smaller in magnitude; compare Figures 6b and S7e), which signifies the weakening of the Walker circulation under climate change[68,78].

Finally, to further verify the consistency of the zonally opposing ITCZ shifts with regional energetics, we evaluate the future EFE shifts over the two sectors (see Table 1 and Figure 6c). Note that the EFE variability has been shown to be linked with the ITCZ variability, not only in the zonal mean[1,9], but also over large longitudinal sectors (the ITCZ – EFE link breaks down only over the western and central Pacific)[10]. For a sector with longitudinal boundaries $\lambda_1$ and $\lambda_2$, the sector-mean position of the EFE (or equivalently of the ITCZ), can be approximated to a first order by meridionally expanding (Taylor series) Equation (2) at the equator[10]:

$$[\varphi_{\mathrm{EFE}}]_{\lambda_1}^{\lambda_2} = -\frac{1}{a} \frac{[v_{\chi_0}]_{\lambda_1}^{\lambda_2}}{[Q_0]_{\lambda_1}^{\lambda_2} - \frac{1}{\lambda_2 - \lambda_1} u_{\chi_0}\big|_{\lambda_1}^{\lambda_2}} \qquad (3)$$

where $[\cdot]_{\lambda_1}^{\lambda_2}$ represents the zonal mean over the sector. Our results show that although CMIP6 models do not predict a robust future EFE shift in the global zonal mean (on the order of $0.2 \pm 0.5°$ N; see Table 1), over the Eurasian sector a robust norward shift is revealed on the order of $0.7 \pm 0.4°$ N, while over the Eastern Pacific and Atlantic sector the EFE shifts to the south by $-1.4 \pm 1.1°$ N. Both these shifts are statistically significant ($p < 0.01$), and they explain 30-40% of the inter-model variance of the projected precipitation change (see Figure 6c).

Overall, the results of this section show that the revealed ITCZ shifts exhibit a robust statistical and physical link with the future changes in the regional energy balance. It can be also concluded that CMIP6 models do exhibit a consensus over the two considered sectors, highlighting opposing ITCZ shifts, opposing changes in $Q$, and opposing EFE shifts. This opposing response of all these quantities and the corresponding models' consensus have been hidden in the zonal-mean analysis of past work.

### 3. Discussion

In this study, the future ITCZ shifts in response to global climate change were explored as a function of longitude and season using climate model simulations. A zonally opposing response of the location of the ITCZ was revealed, which was found to be robust across different climate models, and different seasons, and to be of large longitudinal extent, covering about two thirds of the globe. The opposing ITCZ response can be summarized as a northward shift over the eastern Africa and Indian Ocean and a southward shift over the eastern Pacific, south America and the Atlantic. The revealed response has been masked in the analysis of zonal-mean ITCZ shifts in previous literature, as well as due to the presence of model biases in the present-day climatology of the ITCZ.

We found that the opposing ITCZ response is driven by a positive IOD-like SST pattern over the Indian Ocean, and high SST warming in low latitudes over the eastern Pacific and Atlantic Oceans. This is consistent with the known coupling between tropical SST and precipitation changes. From an atmospheric energetics perspective, our analysis showed that future climate change induces a zonally opposing change in the hemispheric heating of the atmosphere, as a result of the combined effect of radiative and dynamical processes both in the atmosphere and ocean. These included snow and ice-albedo feedbacks, forcing from reductions of the aerosols, cloud radiative effects, OLR cooling, an AMOC weakening, and increases in Southern Ocean heat uptake. Despite differences in model formulations and physics, most models revealed that future changes in the atmospheric energy budget consist of increases in atmospheric heating over Eurasia and cooling over the Southern Ocean, which contrast with atmospheric cooling over the North Atlantic Ocean as a consequence of an AMOC weakening[50,51]. These changes in the regional extratropical atmospheric heating induce an increase in the southward energy transport over the tropics of eastern Africa and Indian Ocean (and an northward shift in the EFE), and an increase in the northward energy transport over the tropical eastern Pacific and Atlantic Oceans (and a southward shift in the EFE), both of which are physically and statistically consistent with the zonally opposing ITCZ response. We note that further analysis based on careful design of idealized climate experiments is needed to determine causality and the relative contribution of extratropical and tropical mechanisms/forcing to the revealed ITCZ shifts in each sector.

Based on our results, we can simultaneously explain anticipated future increases of drought stress in southeastern Africa and Madagascar, intensifying flooding in southern India[72], and greater drought stress in Central America[53] – large hydrological hotspots of global change[87,88] that will affect the livelihood and food security of billions of people.

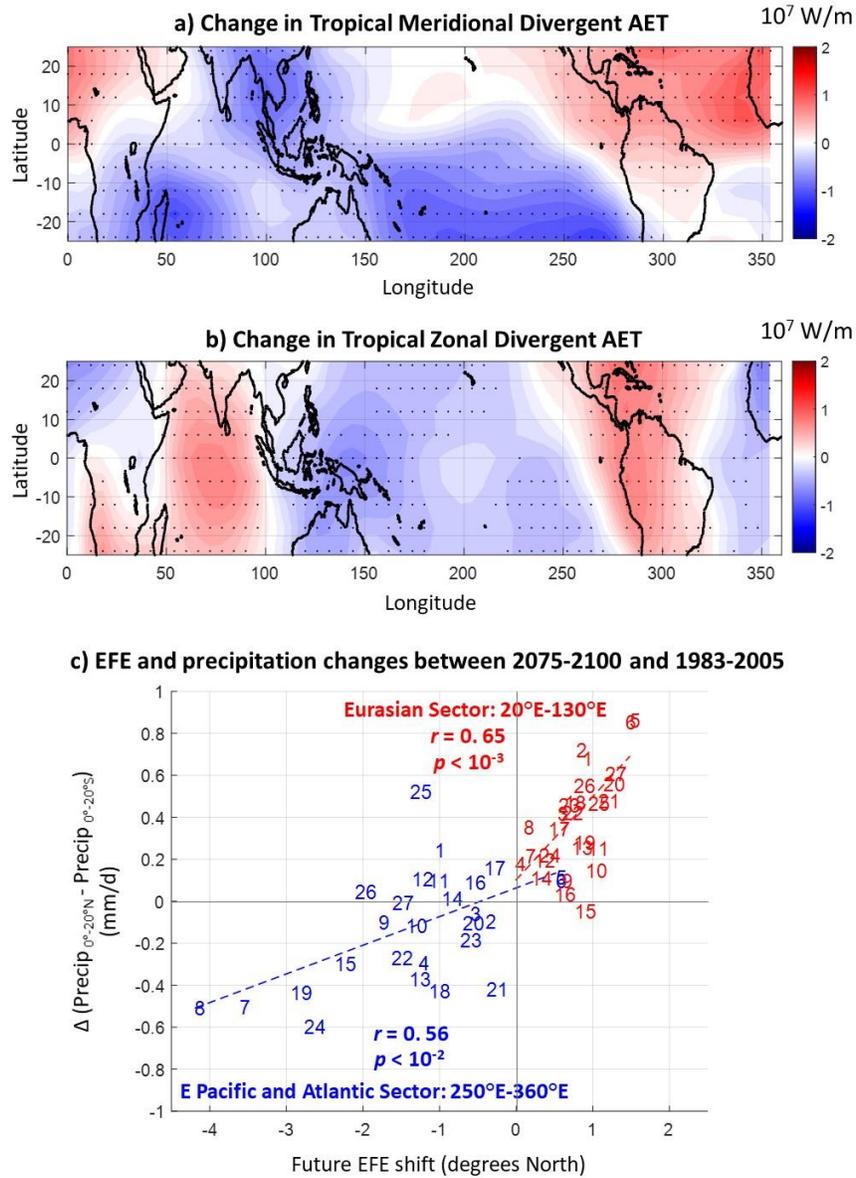

**Figure 6: Future changes in the atmospheric energy transport (AET) over the tropics and the energy flux equator (EFE) in response to climate change, as predicted by CMIP6 models.** a) Change is the divergent meridional component of the atmospheric energy transport over the tropics between 2075-2100 and 1983-2005. The multi-model mean across 27 CMIP6 models is presented, while stippling indicates agreement (in the sign of the change) in more than ¾ of the models considered. b) Same as in (a), but for the divergent zonal component. c) Change in the precipitation asymmetry (between 2075-2100 and 1983-2005) as a function of the EFE shift, using all 27 CMIP6 models zonally averaged over the Eurasian sector (20ºE-130ºE; red color) and the eastern Pacific and Atlantic sector (250ºE-360ºE; blue color). Each model is labeled according to Supplementary Table 1. Results show that under global climate change, the future state of the atmospheric energy transport will be characterized by an increased southward transport (divergent component) over the Eurasian sector, which implies a northward shift of EFE (see Equation 3), and it is statistically consistent with a northward shift of the ITCZ. The opposite (i.e. increased northward energy transport and southward shift of EFE) is true in the eastern Pacific and Atlantic Oceans.

**Methods**

**Probabilistic tracking of the ITCZ.** We seasonally and longitudinally track the ITCZ by simultaneously considering the fields of multiple variables e.g. precipitation, outgoing longwave radiation (OLR), etc., and thus, decreasing the likelihood of detecting spurious ITCZ features that might be detected using only a single variable. Particularly, we use a newly proposed framework which is based on tracking the latitude where the maximum (i.e. above a pre-specified threshold) zonal-mean precipitation and minimum zonal-mean outgoing longwave radiation (OLR) occur in overlapping longitudinal windows[37].

Let $X$ denote the variable (e.g. precipitation) used for defining the ITCZ location, and $X_w^{\lambda,t}$ the zonal mean of $X$ within the longitudinal window [$\lambda$-$w$/2, $\lambda$+$w$/2] of width $w$ and during month/season $t$. The latitudinal distribution of $X_w^{\lambda,t}$ can be obtained from observations or model outputs. For a specified probability of non-exceedance $a$ (tracking threshold), we define $x_{w,a}^{\lambda,t}$ to be the $a$<sup>th</sup> quantile of $X_w^{\lambda,t}$, i.e.,

$$F(x_{w,a}^{\lambda,t}) \equiv \Pr[X_w^{\lambda,t} \leq x_{w,a}^{\lambda,t}] = a$$

where $F$ is the cumulative distribution function (CDF) of $X_w^{\lambda,t}$. We define the random variable $\Phi_{w,a}^{\lambda,t}$ to be the location (in degrees of latitude) at which the ITCZ is most likely to prevail, in longitude $\lambda$, and in month/season $t$. A sample of $\Phi_{w,a}^{\lambda,t}$ may then be the set of latitudinal points $\varphi_{w,a}^{\lambda,t}$ at which the value of $X_w^{\lambda,t}$ exceeds the $a$<sup>th</sup> quantile $x_{w,a}^{\lambda,t}$, that is:

$$\{\varphi_{w,a}^{\lambda,t}\}: X_w^{\lambda,t}(\varphi_{w,a}^{\lambda,t}) > x_{w,a}^{\lambda,t} = F^{-1}(a) \qquad \text{or}$$

$$\{\varphi_{w,a}^{\lambda,t}\}: F\left(X_w^{\lambda,t}(\varphi_{w,a}^{\lambda,t})\right) > a \qquad \text{(M1)}$$

In other words, we track the position of ITCZ based on the upper (1 - $a$)×100% of precipitation in longitude $\lambda$ and month/season $t$, which corresponds to the points $\varphi_{w,a}^{\lambda,t}$. When considering the OLR to track the ITCZ, the negative OLR is used, since deep convection associates with minimum (not maximum) OLR. Such an approach is rather computationally efficient and allows the analysis of both the annual-mean location and the intra-annual variability of the ITCZ, simply by obtaining the ITCZ points, $\varphi_{w,a}^{\lambda,t}$, for each calendar month or each season.

When jointly considering multiple (e.g. $M \geq 2$) variables $\mathbf{X} = [X_1, X_2, \ldots, X_M]$ to track the ITCZ (as in this study), the ITCZ points, $\varphi_{w,a}^{\lambda,t}$, also satisfy Equation (M1), but $F$ is now the joint CDF of $\mathbf{X}_w^{\lambda,t}$.

Herein, we used a non-exceedance $a = 85\%$ as a tracking threshold (general conclusions have been tested across other thresholds too, to ensure robustness), and we averaged precipitation and OLR over longitudinal windows of width $w = 15°$ (see Supplementary Figure 1 for a schematic). However, the framework is general and applicable in considering any single variable, and/or jointly distributed multiple variables to define the ITCZ.

See [37] for more information.

**Definition of ITCZ bias in the models.** The double-ITCZ bias of each CMIP6 model over the eastern Pacific or Atlantic Ocean is defined as the average (over the considered longitudinal sector) difference in the Nov-Apr probability distribution of the ITCZ location between the model and the observations (see Supplementary Figure 2):

$$\Delta P = \frac{1}{\frac{(\lambda_2 - \lambda_1)}{r_\lambda} + 1} \sum_{\lambda=\lambda_1}^{\lambda_2} \left( \frac{1}{2} \int_{\varphi_1}^{\varphi_2} |\Delta \mathrm{PDF}_{\lambda,\varphi}| \, d\varphi \right) \quad (\mathrm{M2})$$

where $\Delta \mathrm{PDF}_{\lambda,\varphi}$ is the difference in the Nov-Apr probability distribution function (PDF) of the ITCZ location between the model and the observations at latitude $\varphi$ and longitude $\lambda$, and $r_\lambda$ is the model's longitudinal resolution. For calculating the bias over the Atlantic ocean, $[\varphi_1, \varphi_2] = [15°S, 10°N]$ and $[\lambda_1, \lambda_2] = [310°E, 360°N]$, while for the eastern Pacific bias, $[\varphi_1, \varphi_2] = [10°S, 15°N]$ and $[\lambda_1, \lambda_2] = [200°E, 300°N]$. The ITCZ biases of all models are presented in Supplementary Table 1. The average bias (weighted by the longitudinal width of each sector) is also presented.

**Correlation significance.** For estimating the $(1 - p)\%$ intervals corresponding to statistically insignificant linear correlation (for a p-value $p$), we assume a $t$-distribution: $r_c = \frac{\pm t}{\sqrt{N-2+t^2}}$, where $t$ is the $(1 - p/2)\%$ quantile of the $t$-distribution, with d.f. = $N$-2, and $N$ is the sample size.

**Data availability.** The data we use in our analysis are all freely available. We use satellite data (monthly precipitation series on a $0.25° \times 0.25°$ grid[59], and OLR series on a $1° \times 1°$ grid[60], for 1983-2005), and climate model outputs from the sixth phase of the Coupled Model Intercomparison Project[54] (CMIP6); see Supplementary Table 1.

**Acknowledgments**

Partial support for this research was provided to EFG, JTR and PS by the National Science Foundation (NSF) under the TRIPODS+ program (DMS-1839336). Moreover, the work of EFG was supported by NSF under the EAGER program (grant ECCS-1839441) and by NASA's Global Precipitation Measurement (GPM) Program (grant NNX16AO56G). JTR received support from DOE's Office of Science RUBISCO Science Focus Area and NASA's SMAP, IDS, and CMS programs. JYY was supported by NSF Climate and Large-scale Dynamics Program of US under Grant AGS-1833075. SY was supported by a generous gift to Yale from Todd Sandoz. A research grant from UCI to advance these research ideas is also acknowledged. We thank the climate modeling groups around the world for producing and making available their model outputs and Dr. Bronwyn Wake (the Editor) and three anonymous reviewers for their constructive comments that helped improve the quality of this paper. We also acknowledge the help from Dr. Ori Adam and Dr. Benjamin Linter in discussing parts of this analysis.

**Author contributions**

AM designed the study, performed the data analysis, and wrote the first draft of the manuscript. All authors contributed to the conceptualization and interpretation of the results and to extended discussions in the revising and finalizing stages of the manuscript.

**Competing interests.** The authors declare no competing financial or non-financial interests.

**Supplementary Material**

Zonally opposing shifts of the intertropical convergence zone in response to climate change by Mamalakis et al.

## Supplementary Discussion

**Climatology of the ITCZ and model biases.** Here, we explore the ability of the considered CMIP6 climate models in accurately reproducing the recent ITCZ climatology and interannual variability. In doing so, we compare the distributions of the location of the ITCZ in May-Oct and Nov-Apr (during the base period 1983-2005) as derived by using satellite observations with those derived by using model outputs (see Figure 1). The analysis of observations indicates that during May-Oct, the ITCZ is a zonally oriented feature located mainly in the Northern Hemisphere, apart from the western Indian Ocean, where it prevails in the southern hemisphere, and the western Pacific, where the tropical part of the south Pacific convergence zone (SPCZ) is also tracked by our method. In the Nov-Apr period, the ITCZ migrates to the south (mainly over land), and three southern convergence features strengthen: the SPCZ (Haffke and Magnusdottir, 2013), the south Atlantic convergence zone (SACZ; Carvalho et al., 2004), and the south Indian ocean convergence zone (SIOCZ; Cook, 1998; 2000), which, in contrast to the summer ITCZ, are diagonally oriented (Widlansky et al., 2011; Barimalala et al., 2018). The highest intra-annual variability of the ITCZ location is found in the western and central Pacific (Mamalakis and Foufoula-Georgiou, 2018), where the ITCZ consists of two distinct and much distant zones, the SPCZ and the northern ITCZ. These two bands coexist for most of the year, with the SPCZ strengthening during boreal winter and the northern ITCZ strengthening during boreal summer (see Figure 2 of Waliser and Gautier, 1993; Widlansky et al., 2011; Berry and Reeder, 2014; Haffke and Magnusdottir, 2013, 2015). The smallest intra-annual variability of the ITCZ location is found in the eastern Pacific and Atlantic oceans, where the ITCZ tends to stay in the northern hemisphere during most of the year; however a double ITCZ may form in the eastern Pacific during boreal spring (see Figure 1c and Adam et al., 2016b; Bischoff and Schneider, 2016; Haffke et al., 2016; Yang and Magnusdottir, 2016).

Although CMIP6 models are mostly consistent in simulating the location of the ITCZ during May-Oct, they exhibit important biases in the Pacific and Atlantic oceans during Nov-Apr (see Figure 1 and Supplementary Figure 2). Particularly, models tend to overestimate the probability of the ITCZ migrating to the southern hemisphere over the eastern Pacific and Atlantic Oceans (Oueslati and Bellon, 2015; Samanta et al., 2019). These biases are well known as double-ITCZ biases and need to be considered when studying future ITCZ trends (Dutheil et al., 2019; Samanta et al., 2019). In doing so, we calculated the average difference in the (Nov-Apr) probability distribution of the ITCZ location between models and observations over specific boxes (see Figure 1d,

Supplementary Figure 2a and section *Methods* for more information), which then allowed us to explore how these biases affect future ITCZ changes (see main text). Our results indicated that CMIP6 models simulate a more frequent seasonal ITCZ migration toward the southern Atlantic than what is observed, by $\Delta P = 57 \pm 17.8\%$ (that is the spatially-averaged difference in probability between models and observations over tropical Atlantic), and likewise in the Pacific toward the southeastern sector of the basin, by $\Delta P = 34 \pm 11.3\%$ (see Supplementary Figure 2b and Supplementary Table 1). These numbers show that the Atlantic bias is more severe, and as such, the signature of the seasonal double-ITCZ biases on annual scales is apparent mainly over the Atlantic and not so much over the eastern Pacific basin (see Figure 1f). Note that when we use the average tropical precipitation and/or OLR difference between models and observations to assess the systematic double-ITCZ biases (i.e. not the probabilistic method), we obtain similar results (see Supplementary Figure 3). Moreover, our analysis shows that there is a statistically significant ($p < 0.05$) positive correlation of eastern Pacific and Atlantic biases across the CMIP6 models on the order of 0.42, which indicates that it is unlikely for a model to exhibit relatively important biases only in one of the two basins. Apart from the double-ITCZ bias, climate models from both projects are also shown to produce a more zonally oriented SPCZ than what observations suggest (see also Oueslati and Bellon, 2015).

To explore the ability of the models to accurately simulate the ITCZ on interannual time scales, we compared the effect of the El Niño – Southern Oscillation (ENSO) on the location of the ITCZ, as determined by satellite data and model outputs (Supplementary Figure 4). Specifically, we calculated the difference in the distribution of the ITCZ location between years corresponding to the four strongest El Niño events and the four strongest La Niña events during the 23-yr base period 1983-2005. In models runs, El Niño and La Niña events do not correspond to the same years with reality, thus, we used the Niño 3.4 index to define ENSO events. Results show that CMIP6 models are mostly consistent in reproducing the effect of ENSO on the location of the ITCZ during Nov-Apr (the period when ENSO typically peaks). Results from both the observations and the models indicate that during El Niño conditions, the ITCZ is displaced more equatorward in the Pacific relative to La Niña conditions, due to the anomalous heating in the tropical Pacific Ocean which favors deep convection (Dai and Wigley, 2000; Berry and Reeder, 2014, Adam et al., 2016b).

**Supplementary Table 1:** CMIP6 models used in this study and their double-ITCZ biases. For models with multiple runs, the average value of bias across all runs is presented.

|    | Model | Number of ensembles | East Pacific double-ITCZ Bias | Atlantic double-ITCZ Bias | Average Bias |
|----|-------|---------------------|-------------------------------|---------------------------|--------------|
| 1  | ACCESS-CM2     | 1  | 0.48 | 0.54 | 0.50 |
| 2  | ACCESS-ESM1-5  | 3  | 0.36 | 0.20 | 0.31 |
| 3  | BCC-CSM2-MR    | 1  | 0.53 | 0.60 | 0.55 |
| 4  | CAMS-CSM1-0    | 2  | 0.53 | 0.85 | 0.64 |
| 5  | CanESM5        | 20 | 0.35 | 0.70 | 0.46 |
| 6  | CanESM5-CanOE  | 3  | 0.36 | 0.71 | 0.47 |
| 7  | CESM2          | 6  | 0.23 | 0.40 | 0.29 |
| 8  | CESM2-WACCM    | 1  | 0.22 | 0.46 | 0.30 |
| 9  | CNRM-CM6-1     | 6  | 0.32 | 0.51 | 0.38 |
| 10 | CNRM-CM6-1-HR  | 1  | 0.28 | 0.50 | 0.35 |
| 11 | CNRM-ESM2-1    | 5  | 0.35 | 0.51 | 0.40 |
| 12 | FGOALS-f3-L    | 1  | 0.22 | 0.37 | 0.27 |
| 13 | FGOALS-g3      | 1  | 0.31 | 0.42 | 0.34 |
| 14 | GFDL-ESM4      | 1  | 0.49 | 0.72 | 0.57 |
| 15 | GISS-E2-1-G    | 1  | 0.53 | 0.78 | 0.61 |
| 16 | INM-CM4-8      | 1  | 0.44 | 0.77 | 0.55 |
| 17 | INM-CM5-0      | 5  | 0.43 | 0.67 | 0.51 |
| 18 | IPSL-CM6A-LR   | 11 | 0.19 | 0.74 | 0.38 |
| 19 | KACE-1-0-G     | 3  | 0.47 | 0.39 | 0.45 |
| 20 | MIROC6         | 3  | 0.16 | 0.38 | 0.24 |
| 21 | MIROC-ES2L     | 1  | 0.22 | 0.53 | 0.32 |
| 22 | MPI-ESM1-2-HR  | 10 | 0.34 | 0.83 | 0.50 |
| 23 | MPI-ESM1-2-LR  | 10 | 0.29 | 0.85 | 0.48 |
| 24 | MRI-ESM2-0     | 1  | 0.26 | 0.58 | 0.36 |
| 25 | NorESM2-LM     | 1  | 0.39 | 0.51 | 0.43 |
| 26 | NorESM2-MM     | 1  | 0.21 | 0.29 | 0.23 |
| 27 | UKESM1-0-LL    | 5  | 0.30 | 0.48 | 0.36 |

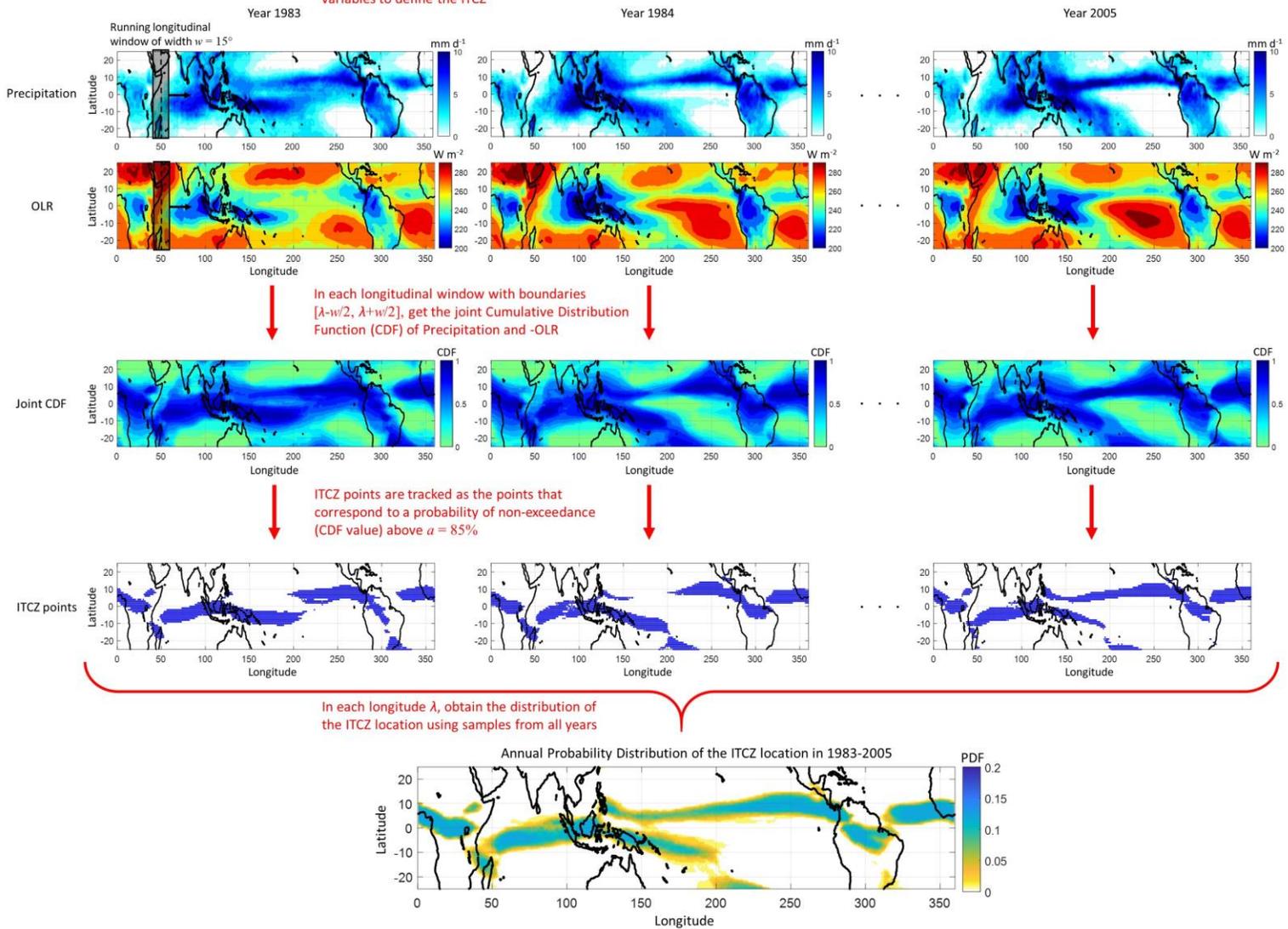

**Supplementary Figure 1: Application of a longitudinally explicit, multivariate probabilistic approach to track the ITCZ on annual scales.** The approach is shown here for the case when the defining variables are $M = 2$, i.e. precipitation and OLR (we use satellite data; see data availability statement), and the tracking probability threshold is $a = 85\%$. When aiming to track the ITCZ on seasonal scales, only the season of interest is used from each year. For more information, please see section *Methods*, and Mamalakis and Foufoula-Georgiou (2018).

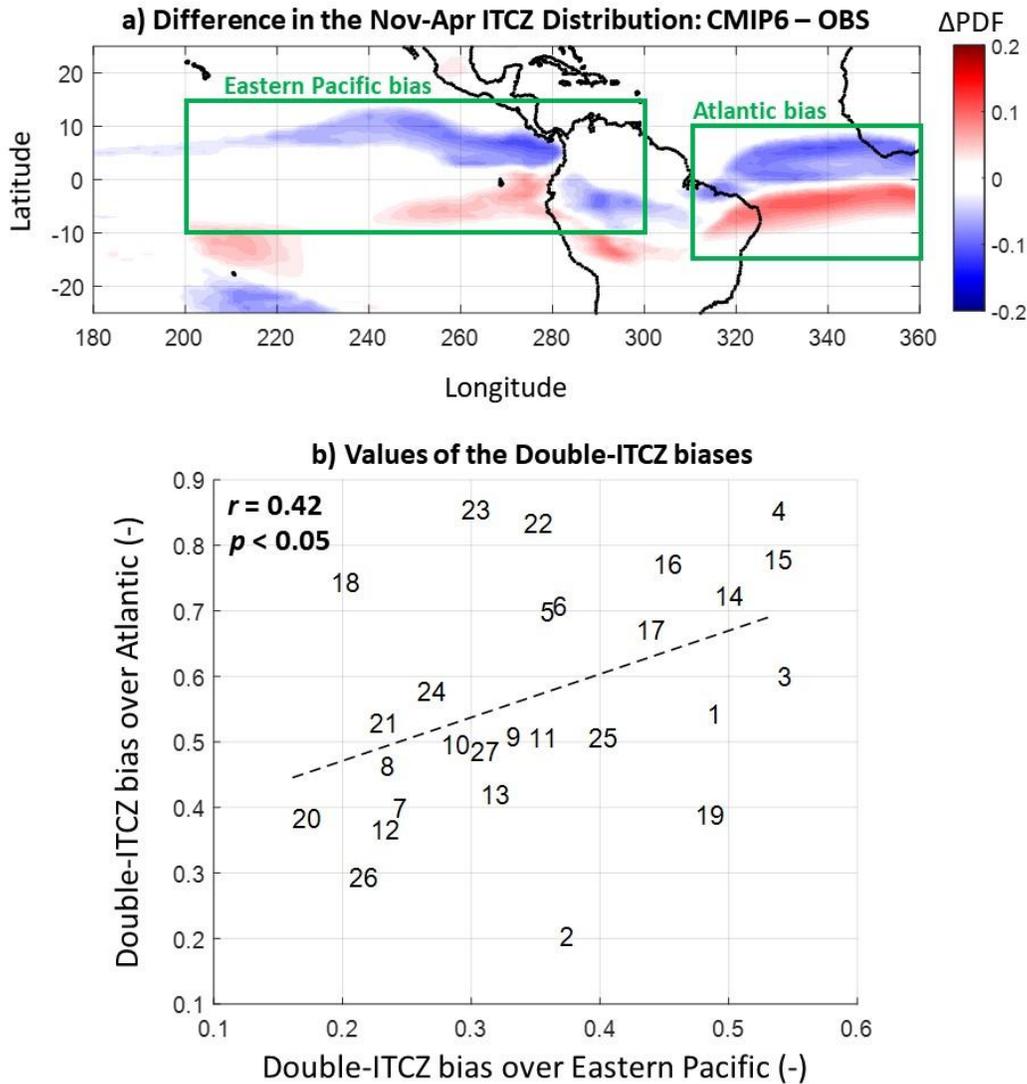

**Supplementary Figure 2: Double-ITCZ biases in CMIP6.** a) Difference in the distribution of ITCZ location (Nov-Apr) between CMIP6 and observations. The multi-model mean across 27 CMIP6 models is presented. b) Scatter plot of the double-ITCZ biases (measured in probability; that is we calculated the average difference in the probability distribution of the ITCZ location between models and observations over the green boxes in Supplementary Figure 2a) in CMIP6 models (each model is labeled according to Supplementary Table 1). The areas over which the double-ITCZ biases are quantified are shown as green boxes in panel (a); see section *Methods* for more information. For models with multiple runs, the average value of bias across all runs is presented. Based on both panels, CMIP6 models are shown to exhibit higher bias over the Atlantic basin than eastern Pacific, while a statistically significant ($p < 0.05$) positive dependence ($r = 0.42$) of these biases is apparent in panel (b). The latter indicates that it is unlikely for a model to exhibit relatively important bias only in one of the two ocean basins.

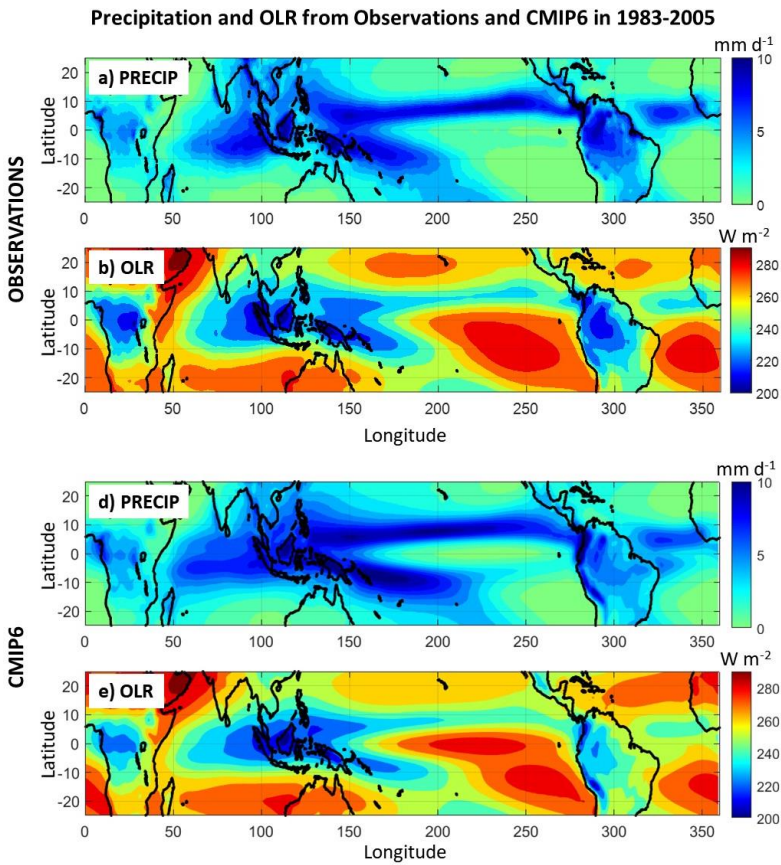
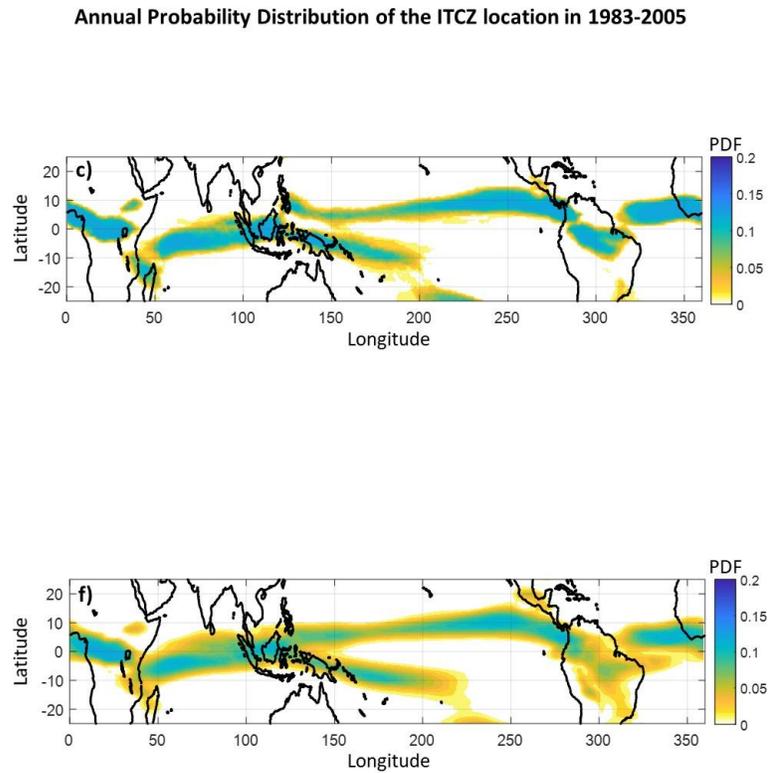

**Supplementary Figure 3: The baseline climatology of the ITCZ in observations and CMIP6, as shown in average precipitation and OLR maps, and using a multivariate probabilistic tracking framework.** a) Observed multi-year mean tropical precipitation in 1983-2005. b) Same as in (a), but for OLR. c) Probability density function (PDF) of the location of the ITCZ on annual scales and in all longitudes. The ITCZ tracking is performed based on the joint statistics of the observed window-mean precipitation and outgoing longwave radiation (OLR) in overlapping longitudinal windows (see Supplementary Figure 1 and section *Methods*); this panel is the identical with Figure 1e. d-f) Same as in (a-c), but results are obtained from the CMIP6 output. The multi-model mean across 27 CMIP6 is presented.

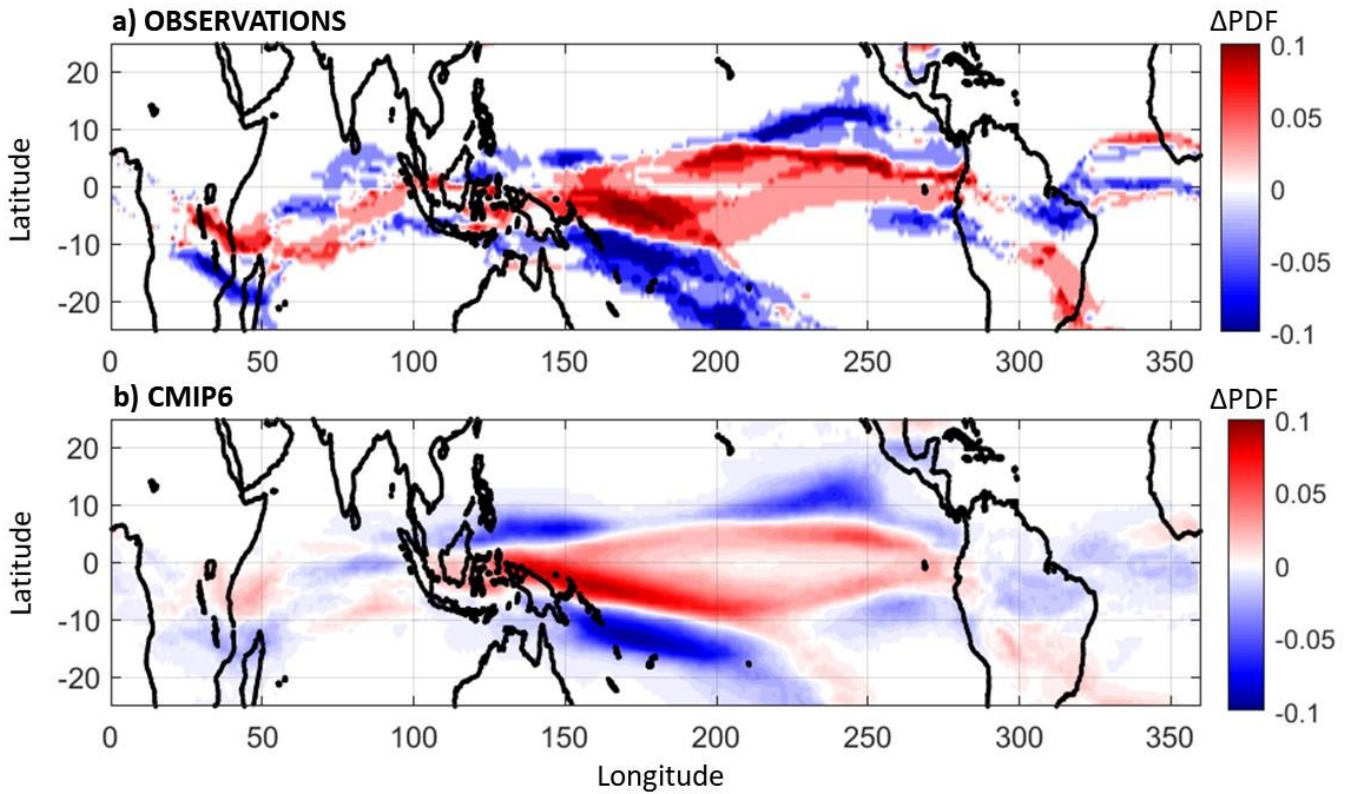

**Supplementary Figure 4: The effect of El Niño-Southern Oscillation on the location of the ITCZ during season Nov-Apr.** a) Difference in the distribution of the ITCZ location between the four strongest El Niño years and the four strongest La Niña years in the observational record in 1983-2005 (the Niño 3.4 index is used to define the El Niño state). It is shown that during El Niño years, the ITCZ is located more equatorward in the Pacific. b) Same as in (a), but the multi-model mean across 27 CMIP6 models is presented, revealing that CMIP6 models capture quite consistently the ENSO effect on the ITCZ.

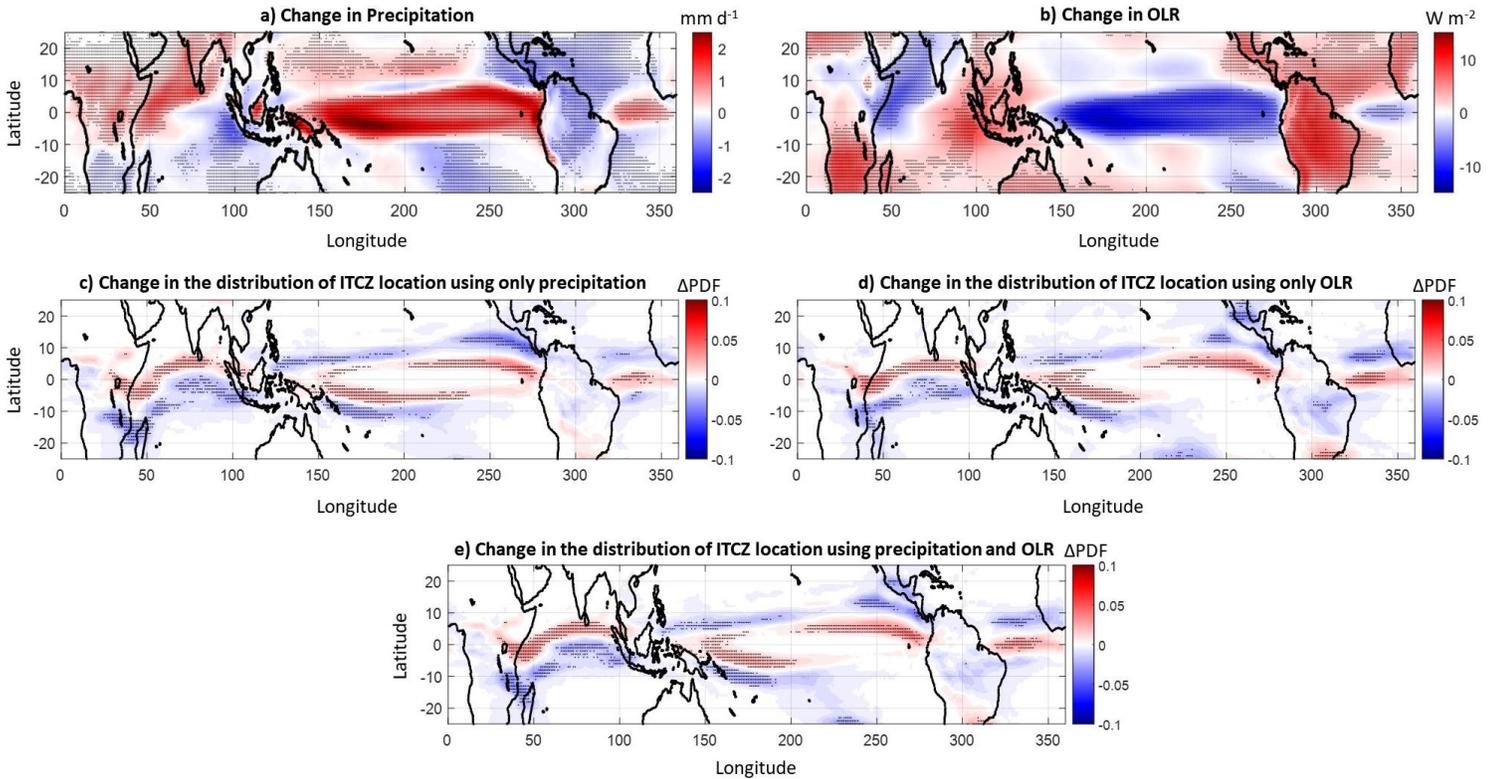

**Supplementary Figure 5: Future changes in the ITCZ location in CMIP6, as shown in changes of average precipitation and OLR maps, and using the multivariate probabilistic tracking framework.** a) Difference of mean precipitation (mm/d) between 2075-2100 and 1983-2005. b) Same as in (a), bur for OLR (W/m$^2$). c) Difference in the annual probability density function (ΔPDF) of the location of the ITCZ between 2075-2100 and 1983-2005. The ITCZ tracking is performed using only precipitation. d) Same as in (c), but OLR is used to track the ITCZ. e) Same as in (c), but both precipitation and OLR are jointly used to track the ITCZ; this panel is the identical with Figure 2c. In all plots, the multi-model mean across 27 CMIP6 models is presented, while stippling indicates agreement (in the sign of the change) in more than ¾ of the models considered. All plots show (to a greater or lesser extend) a northward ITCZ shift over eastern Africa and Indian Ocean and a southward ITCZ shift over eastern Pacific and Atlantic Oceans.

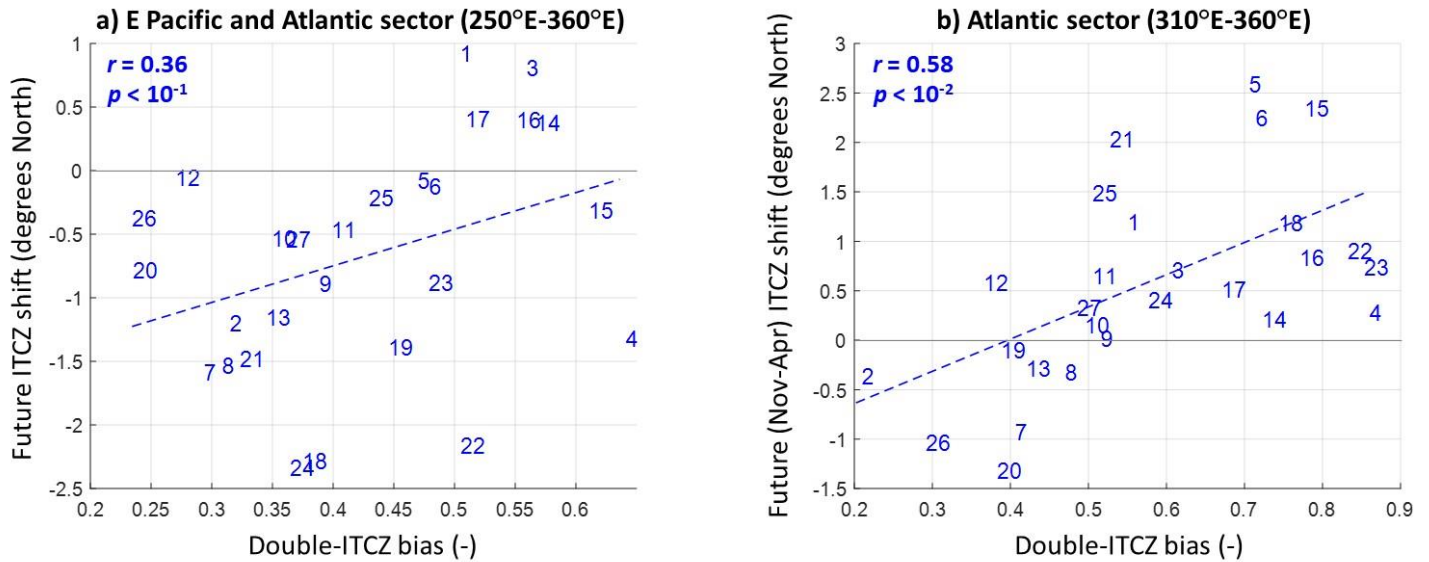

**Supplementary Figure 6: The effect of the double-ITCZ bias on the sign of the projected ITCZ shift over the east Pacific and Atlantic Oceans.** a) The ITCZ shift (in degrees of latitude) over the east Pacific and Atlantic Oceans between 2075-2100 and 1983-2005 is shown as a function of the double-ITCZ bias (measured in probability; that is we calculated the average difference in the probability distribution of the ITCZ location between models and observations over the green boxes in Supplementary Figure 2a) for all CMIP6 models (each model is labeled according to Supplementary Table 1). b) Same as in (a), but results refer to the Atlantic Ocean. In both cases, a statistically significant positive dependence is apparent. This illustrated positive dependence indicates that the lower the double-ITCZ bias of the model over the east Pacific and Atlantic Oceans is in the base period, the more likely it is for the model to produce a southward shift of the ITCZ in the future.

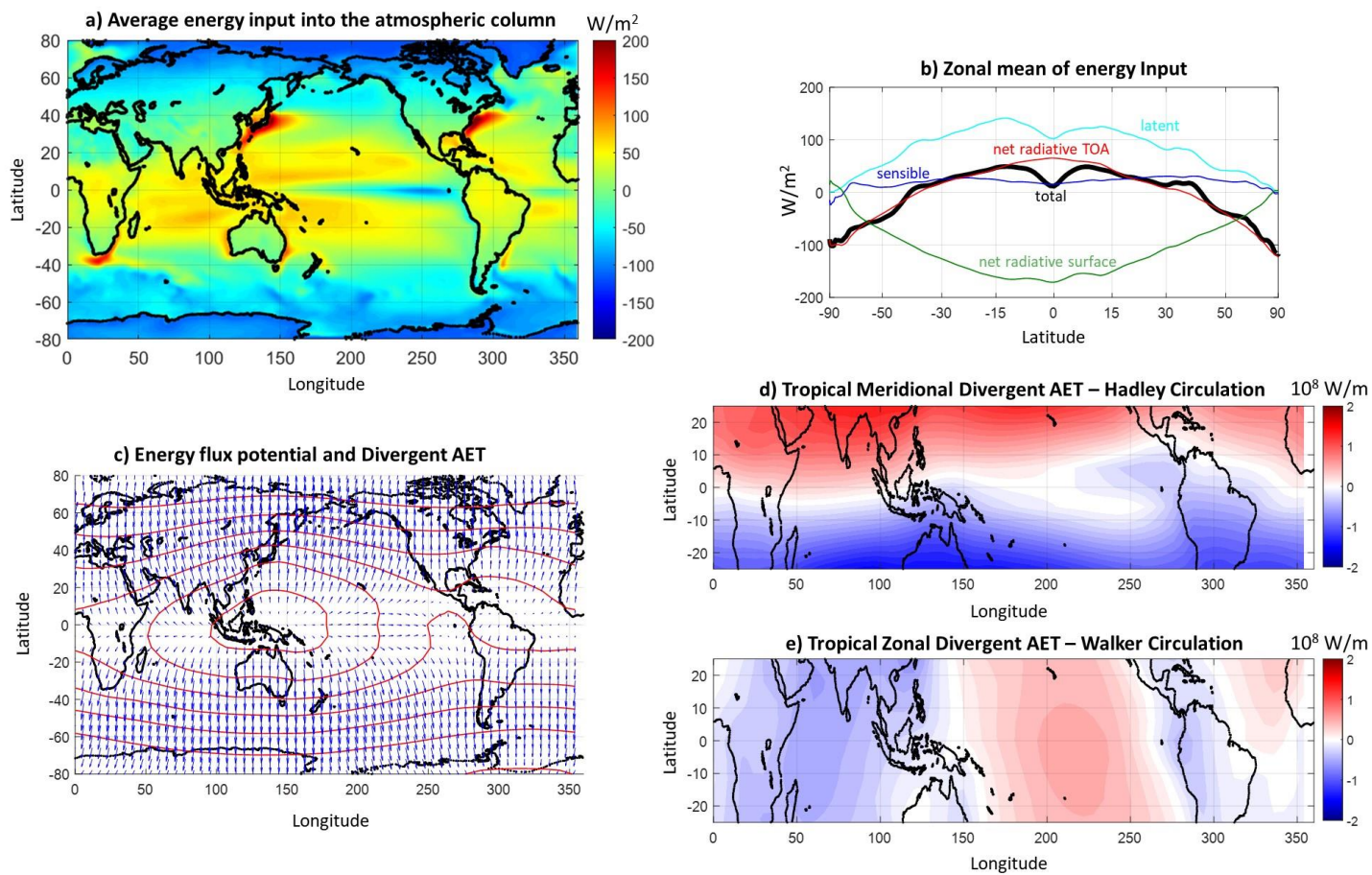

**Supplementary Figure 7: Atmospheric energy transport in base period 1983-2005, as simulated by CMIP6 models.** a) The average energy input (W/m$^2$) into the atmospheric column in the base period 1983-2005. b) Zonal mean of (a). The horizonal axis is scaled as sin($\varphi$). c) Energy flux potential (red contours; contouring interval is 0.2 PW, with equatorial extrema being minima), and divergent atmospheric energy transport (blue vectors). Vectors are on the order of 10$^8$ W/m; see panels (d) and (e) for specific values. d) Divergent meridional component of the atmospheric energy transport over the tropics in 1983-2005, most of which is due to the mean meridional atmospheric circulation (Hadley circulation). e) Same as in (d), but the divergent zonal component is presented (it reflects the Walker circulation). In all plots, the multi-model mean across 27 CMIP6 models is presented.

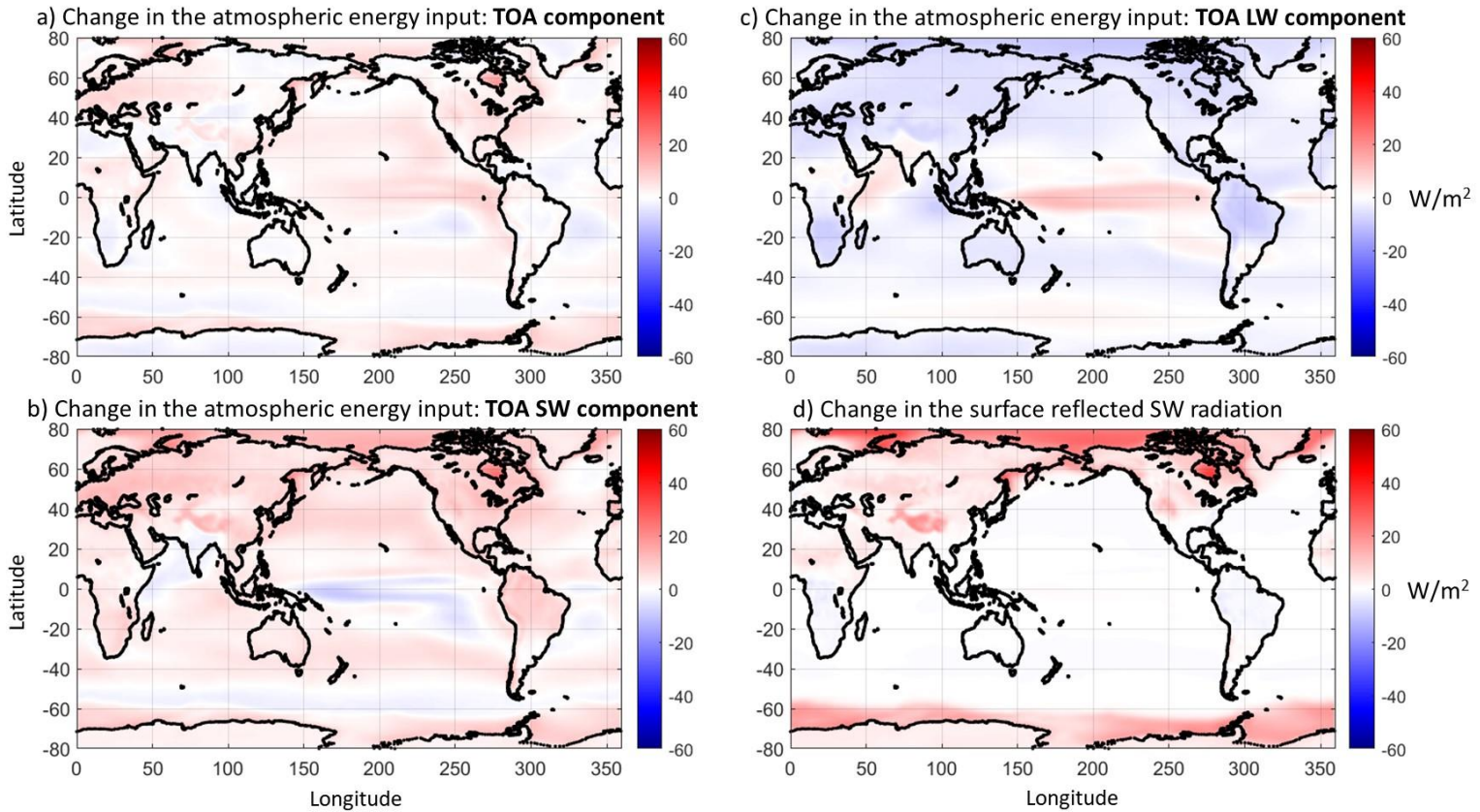

**Supplementary Figure 8: Future changes in the components of the atmospheric energy budget, as predicted in CMIP6.** a) Projected change in the top of the atmosphere (TOA) atmospheric energy input between 2075-2100 and 1983-2005. The multi-model mean across 27 CMIP6 models is presented. b) Same as in (a), but only the TOA shortwave component is presented. c) Same as in (a), but only the TOA longwave component is presented. d) Same as in (a), but the change in the shortwave radiation reaching the surface due to surface albedo changes is presented.